\documentclass[11pt]{article}

\usepackage[dvips]{graphicx}
\usepackage[english]{babel}
\usepackage{psfrag}
\usepackage{slashed}

\usepackage{amsmath}

\usepackage{color,colordvi}

\newcommand{\half}{ {\textstyle\frac{1}{2}} }

\newcommand{\gev}{\mbox{~GeV}}
\newcommand{\mev}{\mbox{~MeV}}

\newcommand{\II}{\hbox{{1}\kern-.25em\hbox{l}}}

\newcommand{\Tr}{\mathop{\rm Tr}\nolimits}
\newcommand \widebar [1] {\overline{#1}}
\newcommand \vev [1] {\langle{#1}\rangle}

\newcommand{\lrD}{{D^{\hspace{-0.8em}%
      \raisebox{0.8ex}{$\scriptstyle\leftrightarrow$}}}{}}
\newcommand{\lD}{{D^{\hspace{-0.8em}%
      \raisebox{0.8ex}{$\scriptstyle\leftarrow$}}}{}}
\newcommand{\rD}{{D^{\hspace{-0.8em}%
      \raisebox{0.8ex}{$\scriptstyle\rightarrow$}}}{}}

\newcommand{\lrpartial}{{\partial^{\hspace{-0.65em}%
      \raisebox{0.8ex}{$\scriptstyle\leftrightarrow$}}}\hspace{-0.05em}{}}
\newcommand{\lpartial}{{\partial^{\hspace{-0.65em}%
      \raisebox{0.8ex}{$\scriptstyle\leftarrow$}}}\hspace{-0.05em}{}}
\newcommand{\rpartial}{{\partial^{\hspace{-0.65em}%
      \raisebox{0.8ex}{$\scriptstyle\rightarrow$}}}\hspace{-0.05em}{}}

\newcommand{\lrnab}{{\nabla^{\hspace{-0.8em}%
      \raisebox{0.8ex}{$\scriptstyle\leftrightarrow$}}}\hspace{-0.05em}{}}
\newcommand{\lnab}{{\nabla^{\hspace{-0.8em}%
      \raisebox{0.8ex}{$\scriptstyle\leftarrow$}}}\hspace{-0.05em}{}}
\newcommand{\rnab}{{\nabla^{\hspace{-0.8em}%
      \raisebox{0.8ex}{$\scriptstyle\rightarrow$}}}\hspace{-0.05em}{}}

\newcommand{\Nvbar}{\widebar{N}_{\!v}}

\newcommand{\slim}{\mskip 1.5mu}

\begin{document}

\begin{titlepage}
\begin{flushright}
\begin{tabular}{l}
DESY-06-123 \\
hep-ph/0608113
\end{tabular}
\end{flushright}

\vspace*{3cm}

\begin{center}

\textbf{\LARGE  Chiral perturbation theory for \\[0.2em]
  nucleon generalized parton distributions}

\vspace*{1.6cm}

{\large
M. Diehl$\slim{}^1$,
A. Manashov$\slim{}^{2,3}$
 and
A.~Sch\"afer$\slim{}^2$
}

\vspace*{0.4cm}

\textsl{%
$^1$ Theory Group, Deutsches Elektronen-Synchroton DESY, 22603
  Hamburg, Germany \\
$^2$ Institut f\"ur Theoretische Physik, Universit\"at Regensburg,
  93040 Regensburg, Germany \\
$^3$ Department of Theoretical Physics, Sankt-Petersburg State
  University, St.-Petersburg, Russia
}

\vspace*{0.8cm}

\textbf{Abstract}\\[10pt]
\parbox[t]{0.9\textwidth}{
We analyze the moments of the isosinglet generalized parton
distributions $H$, $E$, $\tilde H$, $\tilde E$ of the nucleon in
one-loop order of heavy-baryon chiral perturbation theory. We discuss
in detail the construction of the operators in the effective theory
that are required to obtain all corrections to a given order in the
chiral power counting.  The results will serve to improve the
extrapolation of lattice results to the chiral limit.}

\end{center}

\vspace{1cm}

\end{titlepage}


\section{Introduction}
\setcounter{footnote}{0}

In recent years one has learned that many aspects of hadron structure
can be described in the unifying framework of generalized parton
distributions (GPDs).  This framework allows one to combine
information which comes from very different sources in an efficient
and model-independent manner.  The field was pioneered in
\cite{GPDs,Ji,R97} and has evolved to considerable complexity,
reviewed for instance in \cite{Ji98,GPV,Diehl03,Belitsky:2005qn}.  As
GPDs can be analyzed using standard operator product expansion
techniques \cite{GPDs,blumlein}, their moments can be and have been
calculated in lattice QCD \cite{QCDSF}.  Lattice calculations of
well-measured quantities can be used to check the accuracy of the
method, which may then be employed to evaluate quantities that are
much harder to determine experimentally.  This complementarity is
especially valuable in the context of GPDs, because experimental
measurements as e.g.\ in \cite{exp} may not be sufficient to determine
these functions of three kinematic variables in a model-independent
way.  Moreover, several moments of GPDs admit a physically intuitive
interpretation in terms of the spatial and spin structure of hadrons,
see e.g.\ \cite{Ji,Burkardt:2002hr,Polyakov:2002yz,Belitsky:2003nz}.

A notorious problem of lattice QCD is the need for various
extrapolations from the actual simulations with finite lattice
spacing, finite volume and unphysically heavy quarks to the continuum,
infinite volume and physical quark masses.  Simple phenomenological
fits are often still sufficient in view of the general size of
uncertainties, but with increasing numerical precision more reliable
methods have to be applied.  Chiral perturbation theory (ChPT)
provides such a method \cite{ChPT}.  Describing the exact low-energy
limit of QCD it predicts the functional form for the dependence of
observables on the finite volume and the pion mass \cite{chvol} and
also the finite lattice spacing \cite{baer}.  At a given order in the
expansion parameter, ChPT defines a number of low-energy constants
which determine each of these limits.  Some of these constants are
typically known from independent sources, and the remaining ones have
to be determined from fits to the lattice data.  The task of ChPT is
thus to provide the corresponding functional expressions for a
sufficient number of observables.  In this paper we contribute to this
endeavor by analyzing the moments of the isoscalar nucleon GPDs $H$,
$E$, $\tilde H$ and $\tilde E$ in one-loop order.

The analysis of pion GPDs in ChPT has been performed in several
papers~\cite{CJi,KP,DMS}.  In the case of the nucleon GPDs, the chiral
corrections have been calculated for the lowest
moments~\cite{BFHM,CJi,BJi} in the framework of heavy-baryon ChPT,
which performs an expansion in the inverse nucleon mass $1/M$.  Due to
the kinematic limit taken in this scheme, the sum and difference of
the incoming and outgoing nucleon momenta $p^\mu$ and $p'^\mu$ are of
different order in $1/M$.  As a consequence, the $n$th moment of a
nucleon GPD contains terms up to $n$th order in the $1/M$ expansion.
Given the rapidly growing number of low-energy constants in higher
orders of ChPT, it has been assumed that the chiral corrections can
only be calculated for the terms of lowest order in $1/M$, i.e.\ for
the form factors accompanied by the smallest number of vectors
$(p'-p)^\mu$.  This would be a serious setback for the program
sketched above.  The aim of the present paper is to show that the
situation is much better.  In particular, we find that the corrections
of order $O(m_\pi)$ and $O(m_\pi^2)$ to \emph{all} form factors
parameterizing the moments of chiral-even isoscalar nucleon GPDs come
from one-loop diagrams in ChPT and the corresponding higher-order
tree-level insertions.

This paper is organized as follows. In Section~\ref{sec:definitions}
we recall the relation between moments of nucleon GPDs and matrix
elements of twist-two operators and rewrite it in a form suitable for
the $1/M$ expansion.  In Section~\ref{CHPT} we discuss the
construction of twist-two operators in heavy-baryon ChPT and give a
general power-counting scheme for their contribution to a given
nucleon matrix element.  In Sections~\ref{sec:results} and
\ref{sec:gpd-res} we identify the operators that contribute to moments
of GPDs at lowest order in the chiral expansion and give the results
of the corresponding loop calculations.  We summarize our findings in
Section~\ref{sec:sum}.

\section{Generalized parton distributions in the nucleon}
\label{sec:definitions}

The nucleon GPDs can be introduced as matrix elements of nonlocal
operators.  In this paper we limit ourselves to the chiral-even
isoscalar quark GPDs, which are defined by
\begin{align}
  \label{quark-gpd}
\int \frac{d \lambda}{4\pi}\, e^{ix \lambda(aP)}
\langle p'|\, \bar{q}(-\half\lambda a)\, \slashed{a} \,
  q(\half\lambda a) \,|p \rangle
&= \frac{1}{2a P}\, \bar{u}(p') \left[
  \slashed{a}\, H(x,\xi,t) +
  \frac{i \sigma^{\mu\nu} a_{\mu} \Delta_\nu}{2M} \, E(x,\xi,t)\,
\right]  u(p) \,,
\nonumber \\[4mm]
\int \frac{d \lambda}{4\pi}\, e^{ix \lambda(aP)}
  \langle p'|\, \bar{q}(-\half\lambda a)\, \slashed{a}\gamma_5 \,
  q(\half\lambda a) \,|p \rangle
&= \frac{1}{2a P}\, \bar{u}(p') \left[
  \slashed{a}\gamma_5\, \widetilde{H}(x,\xi,t) +
  \frac{a \Delta}{2M}\gamma_5\, \widetilde{E}(x,\xi,t) \right]  u(p) \,,
\hfill
\end{align}
where a sum over $u$ and $d$ quark fields on the l.h.s.\ is
understood, so that $H=H^u+H^d$ etc.  Here $a$~is a light-like
auxiliary vector, $M$ is the nucleon mass, and we use the standard
notations for the kinematical variables
\begin{align}
P=\frac{1}{2} (p+p'),& &\qquad\Delta=p'-p,&
&\qquad t=\Delta^2,& &\qquad \xi=-\frac{\Delta  a}{2P  a} \,.
\end{align}
As usual, Wilson lines between the quark fields are to be inserted in
\eqref{quark-gpd} if one is not working in the light-cone gauge $a^\mu
A_\mu =0$.  The $x$-moments of the nucleon GPDs are related to the
matrix elements of the local twist-two operators
\begin{align}
\label{operators}
\mathcal{O}_{\mu_1 \mu_2 \ldots \mu_n}=
  \mathbf{S}\, \bar{q} \gamma_{\mu_1}
       i \lrD_{\mu_2} \ldots i \lrD_{\mu_n}\, q \, , & &
\widetilde{\mathcal{O}}_{\mu_1 \mu_2 \ldots \mu_n}=
  \mathbf{S}\, \bar{q} \gamma_{\mu_1}\gamma_5\,
       i \lrD_{\mu_2} \ldots i \lrD_{\mu_n}\, q \,,
\end{align}
where $\lrD^\mu=\frac12(\rD^\mu-\lD^\mu)$ and $\mathbf S$ denotes the
symmetrization of all uncontracted Lorentz indices and the subtraction
of traces, e.g.\ $\mathbf{S}\, t_{\mu\nu} = \half (t_{\mu\nu} +
t_{\nu\mu}) - \frac{1}{4}\slim g_{\mu\nu}\slim
t^{\lambda}{}_{\lambda}$ for a tensor of rank two.  It is convenient to
contract all open Lorentz indices with the auxiliary vector $a$,
\begin{align}
\label{defOa}
{\mathcal O}_{\mu_1\ldots\mu_n} ~\to~ {\mathcal O}_n(a)=
a^{\mu_1}\ldots a^{\mu_n}\, {\mathcal O}_{\mu_1\ldots\mu_n}\,,
\end{align}
and in analogy for $\widetilde{\mathcal{O}}$.  The matrix elements of
the operators \eqref{operators} can be parameterized
as~\cite{Ji98,Diehl03}
\begin{align}
  \label{nucl-mat}
\vev{p'|\mathcal{O}_n(a)|p} &=
\sum_{\substack{k=0\\{\mathrm{even}}}}^{n-1}
  (a P)^{n-k-1}\, (a  \Delta)^k\,\,
\bar u(p') \left[ \slashed{a}\, A_{n,k}(t)
           +\frac{i\sigma^{\mu\nu} a_{\mu}\Delta_\nu}{2M}\, B_{n,k}(t)
\right] u(p)
\nonumber \\
&\quad + \bmod(n+1,2)\, (a\Delta)^n\,
         \frac{1}{M}\,\bar u(p') u(p)\, C_n(t)\,,
\nonumber\\[2mm]
\vev{p'|\widetilde{\mathcal{O}}_n(a)|p} &=
\sum_{\substack{k=0\\{\mathrm{even}}}}^{n-1}
  (a P)^{n-k-1}\, (a  \Delta)^k\,\,
\bar u(p') \left[ \slashed{a}\gamma_5\, \widetilde{A}_{n,k}(t)
           +\frac{a \Delta}{2M}\gamma_5\, \widetilde{B}_{n,k}(t)
\right] u(p) \,.
\end{align}
The moments of the above GPDs are polynomials in $\xi^2$,
\begin{align}
  \label{gpd-mom}
\int_{-1}^1dx\, x^{n-1}\, H(x,\xi,t) &=
\sum_{\substack{k=0\\{\mathrm{even}}}}^{n-1}
(2\xi)^k\, A_{n,k}(t) + \mathrm{mod}(n+1,2)\,(2\xi)^{n} C_{n}(t)\,,
\nonumber \\
\int_{-1}^1dx\, x^{n-1}\, E(x,\xi,t) &=
\sum_{\substack{k=0\\{\mathrm{even}}}}^{n-1}
(2\xi)^k\, B_{n,k}(t) - \mathrm{mod}(n+1,2)\,(2\xi)^{n} C_{n}(t)\,,
\displaybreak \nonumber \\
\int_{-1}^1dx\, x^{n-1}\, \widetilde{H}(x,\xi,t) &=
\sum_{\substack{k=0\\{\mathrm{even}}}}^{n-1}
(2\xi)^k\, \widetilde{A}_{n,k}(t)\,,
\nonumber \\
\int_{-1}^1dx\, x^{n-1}\, \widetilde{E}(x,\xi,t) &=
\sum_{\substack{k=0\\{\mathrm{even}}}}^{n-1}
(2\xi)^k\, \widetilde{B}_{n,k}(t)\,.
\end{align}
The restriction to even $k$ in \eqref{nucl-mat} and \eqref{gpd-mom} is
a consequence of time reversal invariance.

To calculate the chiral corrections to the nucleons form factors we
shall use the formalism of heavy-baryon chiral perturbation theory,
which treats the nucleon as an infinitely heavy particle and performs
a corresponding non-relativistic expansion~\cite{JM}. The evaluation
of nucleon form factors in heavy-baryon ChPT is simplified if one
works in the Breit frame~\cite{B92}.  It is defined by the condition
$\vec{P}=0$, so that the incoming and outgoing nucleons have opposite
spatial momenta ${\vec{p}\,}'=-\vec{p}=\vec{\Delta}/2$ and the same
energy $p_0' =p_0^{\phantom{'}} =M\gamma$, where
\begin{equation}
  \gamma = \sqrt{1-\Delta^2/4M^2} \,.
\end{equation}
In the heavy-baryon formalism the baryon has a additional quantum
number, the velocity $v$, which in the Breit frame is $v=(1,0,0,0)$.
The incoming and outgoing nucleon momenta are thus given by $p=M\gamma
v- \Delta/2$ and $p'=M\gamma v+ \Delta/2$.

The Dirac algebra simplifies considerably in the heavy-baryon
formulation. All Dirac bilinears can be expressed in terms of the
velocity $v_\mu$ and the spin operator
\begin{equation}
S_\mu=\frac{i}{2}\gamma_5\sigma_{\mu\nu}\,v^\nu .
\end{equation}
Using that $(v \Delta)= (v S) =0$, one finds in particular
\begin{align}
  \label{D-b}
\widebar u(p') u(p) &= \gamma\,\bar u_v(p')\, u_v(p) \,,
\phantom{\frac{1}{M}}
\nonumber \\[2mm]
\widebar u(p')\gamma_\mu u(p) &=
v_\mu\, \bar u_v(p')\, u_v(p)+\frac{1}{M}\, \bar u_v(p')\,
  [S_\mu,(S\Delta)]\, u_v(p)\,,
\nonumber \\[2mm]
\frac{i}{2M}\, \widebar u(p')\,\sigma_{\mu\nu}\Delta^\nu u(p)
  &= v_\mu \frac{\Delta^2}{4M^2}\, \bar u_v(p')\,u_v(p)
  +\frac{1}{M}\, \bar u_v(p')\,[S_\mu,(S\Delta)]\, u_v(p) \,,
\nonumber \\[2.5mm]
\widebar u(p')\gamma_\mu\gamma_5 u(p)
  &= 2\gamma\, \bar u_v(p')S_\mu u_v(p)+
\frac{\Delta_\mu}{2M^2(1+\gamma)}\, \bar u_v(p')\,(S\Delta)\, u_v(p)\,,
\nonumber \\[2mm]
\widebar u(p')\gamma_5 u(p) &= \frac{1}{M}\,
  \bar u_v(p')\,(S\Delta)\, u_v(p)\,,
\end{align}
where the spinors
\begin{align}
  \label{spinors}
u_v(p)  &= \mathcal{N}^{-1}\, \frac{1+\slashed{v}}{2}\,u(p) , &
u_v(p') &=\mathcal{N}^{-1}\, \frac{1+\slashed{v}}{2}\,u(p')
\end{align}
with
\begin{equation}
  \label{spinor-norm}
\mathcal{N} = \sqrt{\frac{M+vp}{2M}} = \sqrt{\frac{M+v p'}{2M}}
            = \sqrt{\frac{1+\gamma}{2} \rule{0pt}{1.35em}}
\end{equation}
are normalized as $\bar u_v(p,s')\, u_v(p,s) = 2M \delta_{s's}$.
With~\eqref{D-b} one obtains the following representation for the
matrix elements \eqref{nucl-mat} in the Breit frame:
\begin{align}
  \label{OT}
\vev{p'|\mathcal{O}_n(a)|p} &=
\sum_{k=0}^{n} (M\gamma)^{n-k-1}\, (a v)^{n-k}\, (a \Delta)^{k-1}\,
\nonumber \\
& \quad {}\times \bar u_v(p')\, \Big[ (a \Delta)\, E_{n,k}(t)
   +\gamma\,[(a S),(S\Delta)]\, M_{n,k-1}(t) \Big] \,u_v(p) \,,
\nonumber \\[2mm]
\vev{p'|\widetilde{\mathcal{O}}_n(a)|p} &=
\sum_{k=1}^{n} (M\gamma)^{n-k}\, (a v)^{n-k}\, (a \Delta)^{k-1}\,
\nonumber \\
& \quad {}\times \bar u_v(p') \left[
   2\gamma\slim (a S)\,{\widetilde E}_{n,k-1}(t)
   +\frac{(a \Delta) (S\Delta)}{2M^2}\,
    {\widetilde M}_{n,k-1}(t) \right] u_v(p) \,, \hspace{4em}
\end{align}
with
\begin{align}
  \label{ff-trafo}
E_{n,k}(t) &= A_{n,k}(t)+\frac{\Delta^2}{4M^2}B_{n,k}(t)
   \hspace{2em}\text{for~} k<n\,,
& E_{n,n}(t) &= \gamma^2 C_n(t)\,,
\nonumber \\[2mm]
M_{n,k}(t)&=A_{n,k}(t)+B_{n,k}(t)\,,
\nonumber \\[2.5mm]
\widetilde{E}_{n,k}(t)&=\widetilde{A}_{n,k}(t)\,,
\nonumber \\[1mm]
{\widetilde M}_{n,k}(t)&= \frac{1}{1+\gamma}\,\widetilde{A}_{n,k}(t)
+\widetilde{B}_{n,k}\,.
\end{align}
The definition of the $E_n$ and ${\widetilde E}_n$ is conventional but
might be confusing as $E_n$ is not the $n$th moment of $E(x,\xi,t)$
etc.  We nevertheless use this notation, in order to make it easier to
compare our results with those in the literature.  Notice that
according to \eqref{nucl-mat} the terms with $E_{n,k}$ in \eqref{OT}
are only nonzero if $k$ is even, whereas those with $M_{n,k-1}$,
$\smash{\widetilde{E}}_{n,k-1}$ and $\smash{\widetilde{M}}_{n,k-1}$
are only nonzero if $k$ is odd.  We will evaluate these form factors
in heavy-baryon ChPT.  It is straightforward to transform back to the
original form factors using
\begin{align}
  \label{ff-inverse}
A_{n,k}(t) &= \frac{1}{\gamma^2} \left[ E_{n,k}(t)
  - \frac{\Delta^2}{4M^2} M_{n,k}(t) \right] \,,
& B_{n,k}(t) &= \frac{1}{\gamma^2}\, \Big[ M_{n,k}(t)
  - E_{n,k}(t) \Big] \,,
\nonumber \\[1mm]
\widetilde{B}_{n,k}(t) &= \widetilde{M}_{n,k}(t)
  - \frac{1}{1+\gamma}\,\widetilde{E}_{n,k}(t) \,.
\end{align}

\section{Twist-two matrix elements in heavy-baryon ChPT}
\label{CHPT}

Heavy-baryon ChPT combines the techniques of chiral perturbation
theory and of heavy-quark effective field theory~\cite{JM} (for a
detailed review see Ref.~\cite{BKM}). The effective Lagrangian
describes the pion-nucleon interactions in the limit when $m_\pi, q\ll
M$, where $q$ is a generic momentum.  In this situation the velocity
$v$ of the nucleon is preserved in the process. One introduces the
nucleon field with velocity $v$ as~\cite{JM}
\begin{equation}
  N(x)=e^{-iM_0 v x}\bigl(N_v(x)+n_v(x)\bigr)\,,
\end{equation}
where $M_0$ is the nucleon mass in the chiral limit.  The fields
$N_v(x)$, $n_v(x)$ respectively contain the large and small components
of the nucleon field and satisfy $\slashed{v}N_v=N_v$,
$\slashed{v}n_v=-n_v$.  Their Fourier transform depends on the residual
nucleon momentum, i.e.\ the original nucleon momentum minus $M_0 v$.
Integrating out the field $n_v(x)$, one obtains an effective
Lagrangian for the pion-nucleon system which involves the nucleon
field $N_v(x)$ and pion field $\pi(x)$,
\begin{equation}
{\mathcal L}_{\mathrm{eff}}={\mathcal L}_{\pi}+{\mathcal L}_{\pi N}\,,
\end{equation}
where
\begin{align}
{\mathcal L}_{\pi}&={\mathcal L}_{\pi}^{(2)}+{\mathcal L}_{\pi}^{(4)}
  +\ldots\,, &
{\mathcal L}_{\pi N}&={\mathcal L}_{\pi N}^{(1)}
  +{\mathcal L}_{\pi N}^{(2)}+\ldots
\end{align}
are expanded in powers of $q$.  The explicit expressions for the
lowest-order terms read \cite{BKM}
\begin{align}
  \label{L2}
{\mathcal L}^{(2)}_{\pi} &= \frac{F^2}{4}
\Tr\left( \partial_\mu U\partial^\mu U^\dagger
+(\chi^\dagger U+ U^\dagger \chi ) \right)\,,
\nonumber \\[0.4em]
{\mathcal L}^{(1)}_{\pi N} &=
\Nvbar\, \Big\{ i\slim (v \nabla)+g_A (Su) \Big\} \,N_v\,,
\nonumber \\[0.1em]
{\mathcal L}^{(2)}_{\pi N} &= \Nvbar \left\{
  \frac{(v\nabla)^2 -\nabla^2}{2M_0}
  - \frac{ig_A}{2M_0}\, \bigl\{ (\nabla S), (v u)\bigr\}
  + c_1 \Tr\Big( u^\dagger \chi u^\dagger + u \chi^\dagger u \Big) \right.
\nonumber \\
& \hspace{2.2em} \left. {}
  + \Big( c_2 -\frac{g_A^2}{8M_0} \Big) (vu)^2 + c_3\, u_\mu u^\mu
  + \Big( c_4 + \frac{1}{4M_0} \Big) [S^\mu, S^\nu]\, u_\mu u_\nu
  \right\} N_v
\end{align}
with $U=u^2=\exp\{i\pi^a\tau^a/F\}$, the covariant derivative
$\nabla_\mu=\partial_\mu+\Gamma_\mu$, and
\begin{align}
  \label{VA}
\Gamma_\mu&= \frac{1}{2}
 \left(u^\dagger \partial_\mu u+u\partial_\mu u^\dagger\right)
 =\frac{i}{4F^2}\, \epsilon^{abc}\, \pi^a\, \partial_\mu\pi^b \tau^c
  +O(\pi^4) \,,
\nonumber \\[0.2em]
u_\mu&= {i}\left(u^\dagger \partial_\mu u-u\partial_\mu u^\dagger\right)
 \hspace{0.95ex}= -\frac{1}{F}\,\partial_\mu \pi^a\tau^a +O(\pi^3) \,.
\end{align}
The trace $\Tr$ and the Pauli matrices $\tau^a$ refer to isospin
space.  As is done in current lattice QCD calculations, we assume
isospin symmetry to be exact here, neglecting the difference between
$u$- and $d$-quark masses.  The leading-order parameters appearing
in~\eqref{L2} are the pion decay constant $F$ (normalized to $F\approx
92 \mev$) and the nucleon axial-vector coupling $g_A$, both taken in
the chiral limit.  The field $\chi$ implements the explicit breaking
of chiral symmetry by the quark masses, and in the isospin limit can
be replaced by $\chi\to m^2\, \II$, where $m$ is the bare pion and
$\II$ the unit matrix in isospin space.  Estimates of the low-energy
constants $c_i$ in the second-order Lagrangian ${\mathcal
L}^{(2)}_{\pi N}$, which are of order $1/M$, can be found in
\cite{Meissner:2005ba}.  We note that ${\mathcal L}^{(2)}_{\pi N}$
induces corrections to the nucleon propagator, which we treat as
insertions on a nucleon line.  They read $-i\, \bigl( (vl)^2 - l^2
\bigr) \big/(2 M_0)$ and $4i\slim c_1 m^2$, where $l$ is the residual
nucleon momentum, and are to be multiplied with a nucleon propagator
$i/(vl + i\slim 0)$ from ${\mathcal L}^{(1)}_{\pi N}$ on either side.
The pion-nucleon vertices following from ${\mathcal L}^{(2)}_{\pi N}$
can be found in Appendix~A of \cite{BKM}.

In the following subsection we discuss how to construct the operators
in the effective theory that match the twist-two quark operators
\eqref{operators}.  Nucleon matrix elements in the Breit frame are
then obtained as \cite{Steininger:1998ya}
\begin{equation}
  \label{matching}
\vev{p'|\mathcal{O}|p} = \mathcal{N}^2 Z_N\;
  \widebar{u}_v(p')\, G_{\mathcal{O}}(r',r)\, u_v(p) \,,
\end{equation}
with the spinors $u_v$ and normalization $\mathcal{N}$ given in
\eqref{spinors} and \eqref{spinor-norm}.  Here $G_{\mathcal{O}}(r',r)$
is the truncated Green function for external heavy-baryon fields
$\Nvbar$, $N_v$ and the operator $\mathcal{O}$ in the effective
theory.  The residual momenta of the incoming and outgoing nucleon are
given by
\begin{align}
r  &= p  - M_0\slim v = wv - \Delta/2 \,, &
r' &= p' - M_0\slim v = wv + \Delta/2
\end{align}
with
\begin{equation}
  \label{w-def}
w = M (\gamma -1) + \delta M
  = - \frac{\Delta^2}{8M} - 4c_1 m^2 + O(q^3) \,,
\end{equation}
where $\delta M = M-M_0$ is the nucleon mass shift.  Finally, $Z_N$ is
the heavy-baryon field renormalization constant,
\begin{equation}
  \label{ZN}
Z_N = 1 - \frac{3 m^2 g_A^2}{2\slim (4\pi F)^2}
  - \frac{9 m^2 g_A^2}{4\slim (4\pi F)^2}
    \log\frac{m^2}{\mu^2} - 8m^2\, d_{28}^{\slim r}(\mu) + O(q^3) \,,
\end{equation}
where $d_{28}^{\slim r}(\mu)$ is a low-energy constant in the
Lagrangian $\mathcal{L}_{\pi N}^{(3)}$.  As explained in
\cite{Fettes:1998ud} the corresponding operator is required for
renormalization but does not appear in physical matrix elements.  The
value of $d_{28}^{\slim r}(\mu)$ can therefore be chosen freely (with
different choices resulting in different values for other low-energy
constants), and in \cite{Steininger:1998ya} it was chosen such that it
compensates the $\log (m^2/\mu^2)$ term in \eqref{ZN} at the
\emph{physical} value of $m$.  Since we are interested in the pion
mass dependence of matrix elements, we must explicitly keep the
logarithmic term in $Z_N$.  For further discussion we refer to
Section~\ref{sec:tree}.

\subsection{Construction of effective operators}
\label{sec:operators}

We now discuss how to construct the isoscalar local twist-two
operators in the effective theory that match the quark-gluon operators
$\mathcal{O}(a)$ defined in \eqref{operators} and \eqref{defOa}. The
relevant operators in the effective theory can be divided into two
groups: operators ${\mathcal O}_\pi$ which contain only pion fields
(and couple to the nucleon via pion loops) and operators ${\mathcal
O}_{\pi N}$ which are bilinear in the nucleon field.  The matching of
operators thus takes the form
\begin{align}\label{OOa}
{\mathcal O}(a) &\,\cong\,
  {\mathcal O}_\pi(a)+{\mathcal O}_{\pi N}(a)\,, &
\widetilde{\mathcal O}(a) &\,\cong\,
  \widetilde{\mathcal O}_{\pi N}(a) \,,
\end{align}
where we have taken into account that there is no isoscalar pion
operator of negative parity (i.e.\ no $\widetilde{\mathcal
O}_{\pi}(a)$).  The pion isoscalar operators ${\mathcal O}_{\pi}(a)$
have been analyzed in several papers~\cite{ASv,CJi,KP,DMS} and we
shall simply use their results.

Let us now list the building blocks for constructing the operators
${\mathcal O}_{\pi N}(a)$ and $\widetilde{\mathcal O}_{\pi N}(a)$,
which we collectively denote by $Q(a)$, omitting the subscript $\pi N$
for ease of writing.  They should be bilinear in the nucleon field and
should be tensors that have $n$ indices contracted with the auxiliary
vector $a$ according to~\eqref{defOa}.  To build tensors we have the
following objects with Lorentz indices at our disposal: the velocity
vector $v_\mu$, the spin vector $S_\mu$, the derivative
$\partial_\mu$, and the antisymmetric tensor
$\epsilon_{\mu\nu\lambda\rho}$.  We recall that any Dirac matrix
structure can be reduced to an expression containing the spin operator
$S_\mu$, and that the metric tensor $g_{\mu\nu}$ can be omitted in the
construction because the twist-two operators are traceless.  Using the
identities
\begin{align}
\label{SS}
\{S_\mu,S_\nu\} &= \frac{1}{2} ( v_\mu v_\nu-g_{\mu\nu} ) \,, &
[ S_\mu,S_\nu ] &= i\epsilon_{\mu\nu\lambda\rho}\, v^\lambda S^\rho
\end{align}
we can impose that $S_\mu$ should appear at most linearly, or
quadratically as the commutator $[ S_\mu,S_\nu ]$.  Concerning the
derivative $\partial_\mu$, we find it useful to have it acting either
on single nonlinear pion fields $u$, $u^\dagger$ in the combinations
$\Gamma_\mu$ or $u_\mu$ given in \eqref{VA}, or as a total derivative
on the product of all fields, or in the antisymmetric form
$\lrpartial_\mu = \half (\rpartial_\mu - \lpartial_\mu)$ on the
product of all fields to its right or to its left.  This will make it
easy to keep track of factors $\Delta_\mu$ in the corresponding matrix
elements, which play a particular role as we shall see.  To give
operators with the correct chiral transformation behavior, the
derivative $\lrpartial$ must appear in the covariant combination
$\lrnab_\mu = \lrpartial_\mu + \Gamma_\mu$.  The fields and
derivatives used in our construction are then any number of $u_\mu$,
$\lrnab_\mu$ and $\chi_{\pm} = u^\dagger \chi u^\dagger \pm u
\chi^\dagger u$ between the nucleon fields $\Nvbar$ and $N_v$, and any
number of total derivatives $\partial_\mu$ acting on the operator as a
whole.  In the sense of \eqref{VA} we henceforth refer to
$\partial_\mu$, $\lrnab_\mu$ and $u_\mu$ as ``derivatives''.  They
have chiral dimension 1, whereas $\chi_{\pm}$ has chiral dimension 2
and will not appear at the order of the chiral expansion we limit
ourselves to.

We can decompose the pion-nucleon operators $Q_n(a)$ as
\begin{align}
\label{ON}
Q_n(a)= \sum_{k=0}^n M^{n-k-1}\, (a v)^{n-k} Q_{n,k}(a)\,,
\end{align}
where $Q_{n,k}(a) = a_{\mu_1}\ldots a_{\mu_k}\,
Q_{n}^{\mu_1\ldots\mu_k}$ does not contain any factors $(a v)$.  The
$k$ external vectors $a$ in $Q_{n,k}(a)$ can be contracted only with
derivatives $\partial_\mu$, $\lrnab_\mu$, $u_\mu$ and the spin vector
$S_\mu$, or with the antisymmetric tensor.  There can be at most one
factor $(a S)$ as discussed after \eqref{SS}, so that $Q_{n,k}(a)$ has
to contain at least $k-1$ derivatives.  We can hence write\footnote{%
  Instead of $M$ one could also use $M_0$ or $F$ in \protect\eqref{ON}
  and \protect\eqref{ONK}, since all are of the same order in chiral
  power counting.  We find powers of $M$ most convenient, because they
  also appear in the form factor decompositions \protect\eqref{OT}.}
\begin{equation}
  \label{ONK}
Q_{n,k} =M Q_{n,k,-1} +Q_{n,k,0} +\frac{1}{M}\, Q_{n,k,1} +\ldots\,,
\end{equation}
where the operator $Q_{n,k,i}$ has chiral dimension $k+i$.  Note that
due to parity the number of factors $S_{\mu}$, $u_{\mu}$ and
$\epsilon_{\mu\nu\lambda\rho}$ must be even for $\mathcal{O}$ and odd
for $\widetilde{\mathcal{O}}$.  We remark that the contraction of $a$
with the $\epsilon$-tensor involves at least two derivatives, given
that we chose to replace its simultaneous contraction with $v_\lambda$
and $S_\rho$ by $[ S_\mu,S_\nu ]$ using \eqref{SS}.  As a consequence,
the antisymmetric tensor does not appear in the operators with lowest
chiral dimension for a given $k$.

\subsection{Tree-level insertions}
\label{sec:tree}

At tree level, the matrix elements of the effective operators between
two nucleon states are easy to calculate.  Since $u_\mu$ and
$\Gamma_\mu$ involve at least one or two pion fields according to
\eqref{VA}, derivatives in the effective operators are to be replaced
as $\partial_\mu \to i\Delta_\mu$, $u_\mu\to 0$, and $\lrnab_\mu\ \to
-i wv_\mu$ with $w$ given in \eqref{w-def}.  Notice that, while
generically the derivative $\lrnab_\mu$ counts as $O(q)$ in the chiral
expansion, the kinematics of the external nucleon momenta forces $w
v_\mu$ to be of order $O(q^2)$.  As a result, the leading-order
contributions of the operator $Q_{n,k}$ to the form factors in
\eqref{OT} come from the terms with maximum number of factors
$\Delta_\mu$ and no factor $w v_\mu$.  With \eqref{ON} one readily
obtains
\begin{align}
\label{tree}
\vev{p'|\mathcal{O}_{n,k}(a)|p}
  & \,\stackrel{\mathrm{LO}}{=}\,
(a \Delta)^{k-1}\, \bar u_v(p') \left[ (a\Delta)\, E_{n,k}^{(0)}
+ [(a S),(S\Delta)]\, M_{n,k-1}^{(0)} \right] u_v(p)\,,
\nonumber \\[3mm]
\vev{p'|\widetilde{\mathcal{O}}_{n,k}(a)|p}
  & \,\stackrel{\mathrm{LO}}{=}\,
(a\Delta)^{k-1}\, \bar u_v(p')
\left[\slim 2M (a S)\, {\widetilde E}_{n,k-1}^{(0)}
+ \frac{(a\Delta) (S\Delta)}{2M}\,
{\widetilde M}_{n,k-1}^{(0)} \right] u_v(p)\,,
\end{align}
where the superscript on each form factor indicates the term of order
$O(q^0)$ in its chiral expansion.  At this order, the form factors
$E_{n,k}$ and $M_{n,k-1}$ of the vector GPD are related to the matrix
element of the operator $\mathcal{O}_{n,k,0}$, since the nucleon
matrix element of the operator $\mathcal{O}_{n,k,-1}$ is zero at tree
level.  As explained above, this operator contains a factor $(a S)$,
which due to parity must be accompanied by the axial field $u_\mu$ and
hence does not contribute to tree-level matrix elements without
external pions.  For the axial vector GPDs one finds that the form
factor ${\widetilde E}_{n,k-1}$ (${\widetilde M}_{n,k-1}$) receives
its leading contribution from the operator
$\widetilde\mathcal{O}_{n,k,-1}$ ($\widetilde\mathcal{O}_{n,k,1}$),
given the required number of factors $\Delta_\mu$ in \eqref{tree}.

Beyond leading order, tree-level insertions contribute to the form
factors starting at order $O(q^2)$.  Contributions proportional to
$\Delta^2$ are due to operators with $\partial^2$ or to a factor $w$
from operators with $\lrnab$, or to the kinematic factors $\gamma$ in
\eqref{OT} and $\mathcal{N}$ in \eqref{matching}.  Contributions
proportional to $m^2$ are due to operators with $\chi_{+}$ or with
$\lrnab$ and from the wave function renormalization constant $Z_N$ in
\eqref{matching}.  In the results of the following sections we
explicitly include the terms proportional to $g_A^2$ in the expansion
\eqref{ZN} of $Z_N$, whereas contributions from $d_{28}^{\slim r}$ are
lumped into the coefficients describing the $m^2$ corrections due to
tree-level insertions.

\subsection{Loop contributions}
\label{sec:loop}

Let us now consider a loop diagram with the insertion of the operator
$Q_n(a)$.  One easily finds that the term $M^{n-k-1}\,(a v)^{n-k}\,
Q_{n,k}(a)$ in the sum \eqref{ON} can contribute to the form factors
in \eqref{OT} which are accompanied by at least $n-k$ powers of
$(av)$, i.e.\ to $E_{n,m}$, $M_{n,m-1}$, $\widetilde{E}_{n,m-1}$ and
$\widetilde{M}_{n,m-1}$ with $m\le k$.  Chiral counting determines
which terms can contribute to a given order.  Namely, the contribution
of the operator $Q_{n,k,i}$ in a loop diagram has chiral dimension
\begin{equation}
D_{k,i} =4L +(k+i) +\sum_{j=1}^{N_{\pi}} \dim V_{\pi}(j)
 +\sum_{j=1}^{N_{\pi N}}  \dim V_{\pi N}(j) -2 I_{\pi} -I_N\,,
\end{equation}
where $L$ is the number of loops and $(k+i)$ is the chiral dimension
of the operator insertion.  $V_{\pi}(j)$ and $V_{\pi N}(j)$
respectively denote the $j$th vertex from $\mathcal{L}_{\pi}$ and
$\mathcal{L}_{\pi N}$ in the graph.  $N_{\pi}$ and $N_{\pi N}$ are the
corresponding total numbers of vertices, and $I_{\pi}$ and $I_N$ are
the numbers of pion and nucleon propagators.\footnote{%
  Note that a nucleon propagator correction from a higher-order
  Lagrangian counts as one (nucleon-nucleon) vertex with two nucleon
  propagators on either side, see the discussion after
  \protect\eqref{VA}.}
Using the relation $L=I_{\pi}+I_N-N_{\pi}-N_{\pi N}$ (see e.g.\
\cite{BKM}) and the fact that for our specific diagrams $I_N=N_{\pi
N}$, we can rewrite this expression as a sum of positive terms, which
makes it easy to identify the different contributions at a given
order:
\begin{equation}
  \label{pow-count}
D_{k,i}= 2L-1 +k+(i+1)
 +\sum_{j=1}^{N_{\pi}}\, \bigl(\dim{V_{\pi}}(j)-2 \bigr)
 +\sum_{j=1}^{N_{\pi N}}\, \bigl(\dim V_{\pi N}(j)-1 \bigr)\,.
\end{equation}
For each vertex we can insert either the lowest or any higher order,
i.e.\ $\dim V_{\pi}(j)=2,4,...$ and $\dim V_{\pi N}(j)=1,2,...$.  Note
that a loop diagram with chiral dimension $D_{k,i}$ generates
contributions to a nucleon matrix element of order $O(q^{\slim d})$
with $d\ge D_{k,i}$.  This is on one hand because of the explicit
factors $\mathcal{N}$ and $Z_N$ in \eqref{matching}, and on the other
hand because the sum $r^\mu+r'^\mu = 2w v^\mu$ is of order $O(q^2)$
and thus one order higher than the generic power associated with
residual nucleon momenta.

The form factors enter a matrix element multiplied by factors
$(a\Delta)$ or $(S\Delta)$ as given in \eqref{OT}.  Taking these into
account, one finds that the chiral correction from $Q_{n,k,i}$ to
$E_{n,m}$ and $M_{n,m-1}$ has at least order $D_{k,i}-m$, while for
the form factors $\widetilde{E}_{n,m-1}$ and $\widetilde{M}_{n,m-1}$
it has at least order $D_{k,i}-m+1$ and $D_{k,i}-m-1$, respectively.
This is a main result of our paper and allows one to determine which
operators need to be considered to calculate the corrections to a form
factor to a given order in the chiral expansion.  Because $D_{k,i}$
contains a term $k$ and because of the constraint $k-m \ge 0$, the
number of loops and the order of the chiral Lagrangian required to
calculate the lowest-order corrections for a given form factor do
\emph{not} grow with $m$.  Instead, a growing number of factors
$\Delta_\mu$ accompanying a form factor in the nucleon matrix element
requires a growing number of derivatives in the operator $Q_{n,k}$.

As an application of our general result we find that the form factors
$E_{n,m}$ and $M_{n,m-1}$ can receive corrections starting at order
\begin{itemize}
\item $O(q)$ from one-loop diagrams with insertion of the operator
  $Q_{n,m,-1}$ and leading-order (LO) pion-nucleon vertices,
\item $O(q^2)$ from the one-loop diagrams with insertion of the
  operators $Q_{n,m,0}$ and $Q_{n,m+1,-1}$ and LO pion-nucleon
  vertices,
  and from one-loop diagrams with insertion of the operator
  $Q_{n,m,-1}$ and one next-to-leading order (NLO) pion-nucleon vertex
  or nucleon propagator correction.
\end{itemize}
In turn, the form factor $\widetilde{E}_{n,m-1}$ receives corrections
starting at order $O(q^2)$ from one-loop diagrams with leading-order
vertices and insertion of the operator $Q_{n,m,-1}$.  For
$\widetilde{M}_{n,m-1}$ the discussion of corrections up to order
$O(q^2)$ is more involved and will be given in
Section~\ref{sec:axial-results}.

To conclude the discussion of power counting, we consider the
contribution to the form factors $E_{n,m}$ and $M_{n,m-1}$ of loop
graphs with the insertion of the pion operators $\mathcal{O}_\pi(a)$,
see~\eqref{OOa}.  Repeating the above argument and taking into account
that now $I_N = N_{\pi N} -1$, one finds that such diagrams have
chiral dimension
\begin{equation}
D_\pi= 2L-1 +\dim{\mathcal{O}_{\pi}}
+\sum_{j=1}^{N_{\pi}}\, \bigl(\dim{V_{\pi}(j)}-2 \bigr)
+\sum_{j=1}^{N_{\pi N}}\, \bigl(\dim V_{\pi N}(j)-1 \bigr)\,.
\end{equation}
Given that the leading operator $\mathcal{O}_{\pi}^{n}(a)$
contributing to $\mathcal{O}_n(a)$ has the chiral dimension $n$, one
finds that it can contribute to the form factors $E_{n,m}$ and
$M_{n,m-1}$ starting at order $O(q^{n-m+1})$.  Note that because of
charge conjugation invariance the isoscalar pion operators
$\mathcal{O}_{\pi}^{n}(a)$ have even $n$ and that due to time reversal
invariance the form factors $E_{n,k}$ and $M_{n,k}$ vanish for odd
$k$.  Together with our power-counting formula one thus finds that
$E_{n,n}$ gets contributions from $\mathcal{O}_{\pi}^{n}(a)$ starting
at order $O(q)$ and $M_{n,n-2}$ starting at order $O(q^2)$.  All other
corrections from operators $\mathcal{O}_\pi(a)$ to form factors
$E_{n,k}$ and $M_{n,k}$ start at $O(q^3)$.

\begin{figure}[t]
\psfrag{l}[c][c][0.8]{$l$}
\psfrag{a}[c][c][0.8]{$wv-\frac{\Delta}2$}
\psfrag{c}[c][c][0.8]{$wv+\frac{\Delta}2$}
\psfrag{e}[c][c][0.8]{$l+wv-\frac{\Delta}2$}
\psfrag{o}[c][c][0.8]{$l+wv+\frac{\Delta}2$}
\psfrag{aa}[b][b][1.0]{a}
\psfrag{bb}[b][b][1.0]{b}
\psfrag{cc}[b][b][1.0]{c}
\centerline{\includegraphics[width=16cm]{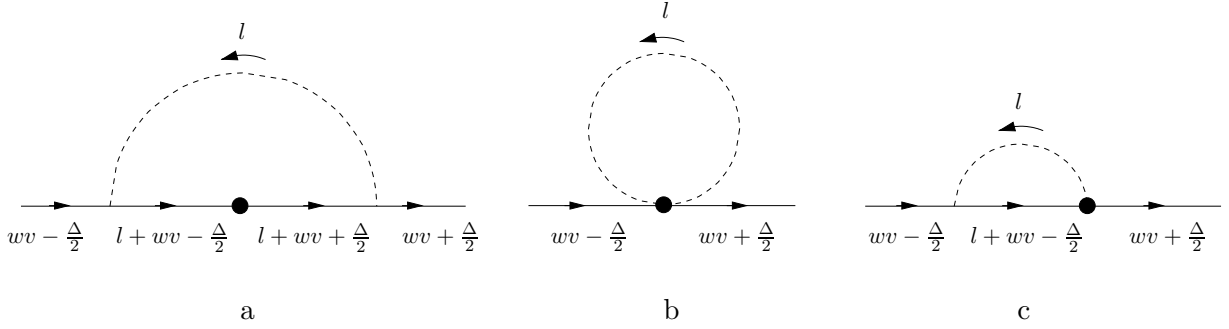}}
\caption{\label{fig-1} One-loop graphs with the insertion of a
pion-nucleon operator $\mathcal{O}_{\pi N}(a)$ or
$\widetilde{\mathcal{O}}_{\pi N}(a)$, denoted by a black blob.  Not
shown is the analog of graph c with residual momentum $l+wv +\Delta/2$
of the intermediate nucleon line.}
\end{figure}

Let us now take a closer look at the one-loop graphs with pion-nucleon
operator insertions, which are shown in Fig.~\ref{fig-1}.  With our
construction of operators explained in Section~\ref{sec:operators} we
can readily analyze the origin of factors $\Delta_\mu$, whose number
determines to which form factor a graph will contribute.  Using $(v
\Delta)= (v S) =0$ and the form \eqref{L2} of the LO and NLO
pion-nucleon Lagrangian, we find that the numerators of the loop
integrals are composed as specified in Table~\ref{tab:vectors}.  The
denominators of the pion and nucleon propagators respectively are
$(l^2 - m^2 +i\slim 0)$ and $(lv +w +i\slim 0)$, so that the loop
integration turns tensors $l_{\mu_1} \ldots l_{\mu_j}$ into tensors
constructed from $v_{\mu}$ and $g_{\mu\nu}$.  A factor $\Delta_\mu$
that can be contracted with $a^\mu$ or $S^\mu$ (i.e.\ is not
contracted to $\Delta^2$) can hence only originate from total
derivatives $\partial_\mu$ in the operator insertion and from an NLO
pion-nucleon vertex or nucleon propagator correction.  We will see
that this reduces considerably the number of operators contributing to
the leading chiral corrections of nucleon GPDs.

\begin{table}
\caption{\label{tab:vectors} Four-vectors and their products appearing
  in the numerators of the loop graphs of Fig.~\protect\ref{fig-1}.
  $NN$ vertices (arising from nucleon propagator corrections) are not
  explicitly shown in the graphs.}
\begin{center}
  \renewcommand{\arraystretch}{1.2}
  \begin{tabular}{lll} \hline\hline
derivatives in operator insertion & $\partial_\mu$ & $\Delta_\mu$ \\
 & $\lrnab_\mu$ & $l_\mu$ ~and~ $w v_\mu$ \\
 & $u_\mu$ & $l_\mu$ \\ \hline
vertices & $\pi NN$ at LO & $Sl$ \\
 & $\pi NN$ at NLO & $(vl) (Sl) \pm (vl) (S\Delta)$ \\
 & $NN$ at NLO & $(vl)^2 -l^2 \pm l\Delta -\Delta^2/4$ \\
\hline\hline
  \end{tabular}
\end{center}
\end{table}

\section{Chiral corrections up to order $O(q^2)$}
\label{sec:results}

\subsection{Axial-vector operators}
\label{sec:axial-results}

Using the formalism developed in the previous section, we now evaluate
the form factors up to relative order $O(q^2)$.  Let us start by
giving the operators in $\widetilde{O}_{n,k,i}$ that have the maximum
number of total derivatives $\partial_\mu$ contracted with $a^\mu$ or
$S^\mu$.  It will turn out that these are required to produce the
factors of $(a \Delta)$ and $(S \Delta)$ in the form factor
decomposition \eqref{OT}.  With the constraints of parity invariance,
we find
\begin{align}
\label{Oa-1}
\widetilde\mathcal{O}_{n,k,-1}(a) &= \tilde{b}_{n,k}\,
  (ia\partial)^{k-1}\, \Nvbar (a S)N_v + \ldots \,,
\nonumber \\[0.3em]
\widetilde\mathcal{O}_{n,k,1}(a) &= \tilde{c}_{n,k}\,
  (ia\partial)^k (i\partial_\mu)\, \Nvbar\, S^\mu N_v + \ldots \,,
\end{align}
where the $\ldots$ stand for operators with fewer total derivatives.
One has ${\widetilde E}_{n,k-1}^{(0)} = \tilde{b}_{n,k}/2$ and
${\widetilde M}_{n,k-1}^{(0)} = 2\tilde{c}_{n,k}$ for the tree-level
contributions at order $O(q^0)$.  {}From the time-reversal constraints
on the form factors it follows that the low-energy constants
$\tilde{b}_{n,k}$ and $\tilde{c}_{n,k}$ are zero for even $k$.

As derived in Section~\ref{sec:loop}, the leading chiral corrections
to $\widetilde{E}_{n,k-1}$ come from one-loop graphs with LO
pion-nucleon vertices and the operator $\widetilde{O}_{n,k,-1}$.
Since this operator does not contain pion fields, one needs to
calculate only graph a in Fig.~\ref{fig-1}.  One finds
\begin{equation}
  \label{res-Et}
\widetilde E_{n,k}(t)=\widetilde
E_{n,k}^{(0)}\left\{1-\frac{3m^2g_A^2}{(4\pi F)^2}
\left[\log\frac{m^2}{\mu^2}+1\right] \right\}
+\widetilde{E}_{n,k}^{(2,m)} m^2 +\widetilde{E}_{n,k}^{(2,t)} t
+O(q^3)\,,
\end{equation}
where the terms going with $m^2$ and $t$ originate from tree-level
insertions as discussed at the end of Section~\ref{sec:tree}.  Here
and in the following we use the subtraction scheme of \cite{ChPT} for
the loop graphs, subtracting $1/\epsilon+\log(4\pi)+\psi(2)$ for each
$1/\epsilon$ pole in $4-2\epsilon$ dimensions.  The renormalization
scale is denoted by $\mu$, and the $\mu$ dependence of the logarithm
in \eqref{res-Et} cancels against the $\mu$ dependence of
$\widetilde{E}_{n,k}^{(2,m)}$, which we have not displayed for
simplicity.  Note that the bare parameters $m$, $F$, $g_A$ can be
replaced with their counterparts at the physical point within the
precision of our result.  Since the nonanalytic corrections in
\eqref{res-Et} are independent of the moment indices $n$ and $k$, they
apply to the entire nucleon GPD $\widetilde H(x,\xi,t)$,
\begin{align}\label{H1loop}
\widetilde H(x,\xi,t)=
\widetilde H^{(0)}(x,\xi)\left\{1-\frac{3m^2g_A^2}{(4\pi F)^2}
\left[\log\frac{m^2}{\mu^2}+1\right] \right\}
+m^2\slim\widetilde{H}^{(2,m)}(x,\xi)
+t\slim \widetilde{H}^{(2,t)}(x,\xi) +O(q^3) \,.
\end{align}

Let us now consider the chiral corrections for $\widetilde M_{n,k-1}$.
It follows from \eqref{OT} that the relevant diagrams have to produce
a factor $(a\Delta)^k (S \Delta)$.  By power counting, the form factor
$\widetilde M_{n,k-1}$ could receive corrections of order $O(q^0)$
from diagrams with LO vertices and the operator insertion
$\widetilde\mathcal{O}_{n,k,-1}$.  Similarly, corrections of order
$O(q)$ could come from the diagrams with LO vertices and insertion of
$\widetilde\mathcal{O}_{n,k+1,-1}$ or $\widetilde\mathcal{O}_{n,k,0}$,
and from diagrams with insertion of $\widetilde\mathcal{O}_{n,k,-1}$
and one NLO pion-nucleon vertex or nucleon propagator correction.  One
finds no operator in $\widetilde \mathcal{O}_{n,k,0}$ that has $k$ or
more partial derivatives contracted with $a^\mu$ or $S^\mu$, and the
same holds of course for $\widetilde \mathcal{O}_{n,k,-1}$.  According
to our discussion in Section~\ref{sec:loop} the graphs just discussed
can thus produce at most $k$ vectors $\Delta_\mu$ (not counting those
appearing in $\Delta^2$) and hence do not contribute to
$\widetilde{M}_{n,k-1}$.  At order $O(q^2)$ there is a number of
possibilities:
\begin{enumerate}
\item graphs with LO vertices and insertion of
  $\widetilde\mathcal{O}_{n,k+2,-1}$,
  $\widetilde\mathcal{O}_{n,k+1,0}$ or
  $\widetilde\mathcal{O}_{n,k,1}$.  The insertion of
  $\widetilde\mathcal{O}_{n,k+1,0}$ does not produce a sufficient
  number of factors $\Delta_\mu$, whereas insertion of
  $\widetilde\mathcal{O}_{n,k+2,-1}$ gives a factor $(a\Delta)^{k+1}
  (aS)$, which contributes to the form factor $\widetilde{E}_{n,k+1}$
  but not to $\widetilde{M}_{n,k-1}$.  A correction to
  $\widetilde{M}_{n,k-1}$ is obtained from insertion of the operator
  $\widetilde\mathcal{O}_{n,k,1}$ given in \eqref{Oa-1}, which already
  provides the tree-level term of this form factor.  Only the loop
  graph in Fig.~\ref{fig-1}a is nonzero for this insertion, and the
  result is analogous to the one for the contribution of
  $\widetilde\mathcal{O}_{n,k,-1}$ to $\widetilde{E}_{n,k-1}$.
\item graphs with one NLO vertex or propagator correction and
  insertion of $\widetilde\mathcal{O}_{n,k+1,-1}$ or
  $\widetilde\mathcal{O}_{n,k,0}$.  Insertion of
  $\widetilde\mathcal{O}_{n,k,0}$ does again not provide enough
  factors of $\Delta_\mu$, whereas graphs with
  $\widetilde\mathcal{O}_{n,k+1,-1}$ give zero due to time reversal
  invariance.  This can be seen by direct calculation, or by noting
  that $\widetilde{M}_{n,k-1}$ is only nonzero for odd $k$, whereas
  the coefficient $\tilde{b}_{n,k+1}$ is only nonzero for even $k$, as
  remarked below \eqref{Oa-1}.
\item graphs with insertion of $\widetilde\mathcal{O}_{n,k,-1}$ and
  $(i)$ two loops with LO vertices, or $(ii)$ one loop with two NLO
  pion-nucleon vertices or nucleon propagator corrections, or $(iii)$
  one loop with one NNLO pion-nucleon vertex or nucleon propagator
  correction, or $(iv)$ one loop with a pion propagator correction
  from $\mathcal{L}_{\pi}^{(4)}$.  The operator insertion provides
  $k-1$ factors of $\Delta_\mu$, so that two more factors must be
  provided by the vertices or propagator corrections (without being
  contracted to $\Delta^2$).  This is not possible in case $(i)$,
  because the LO pion-nucleon vertices only involve pion momenta and
  the pion momenta in a two-loop graph can be parameterized such that
  they are independent of $\Delta$ (as in the one-loop graphs of
  Fig.~\ref{fig-1}).  Likewise, a pion propagator correction in case
  $(iv)$ does not depend on $\Delta$ and can therefore not contribute.
  In cases $(ii)$ and $(iii)$ one obtains nonzero contributions from
  the graph in Fig.~\ref{fig-1}a.  The NNLO vertices and propagator
  corrections follow from the Lagrangian $\mathcal{L}_{\pi N}^{(3)}$
  given in \cite{Fettes:1998ud}.  We find that the only term providing
  the two required factors of $\Delta_\mu$ is the $\pi NN$ vertex
  generated by
  \begin{equation}
    -\frac{g_A}{4M_0^2}\, \Nvbar\, \Big\{
    (\lnab S) (u \rnab) + (\lnab u) (S \rnab) \Big\}\, N_v \,.
  \end{equation}
  Note that this vertex does not introduce a new low-energy constant,
  similarly to the term proportional to $g_A$ in $\mathcal{L}_{\pi
  N}^{(2)}$, which generates the $\pi NN$ coupling at NLO.  These
  terms arise from the $1/M_0$ expansion of the leading-order
  relativistic pion-nucleon Lagrangian $\widebar{N}\,
  (i\slashed{\nabla} - M_0 +\half g_A \slashed{u} \gamma_5)\, N$, see
  e.g.~\cite{BKM}.
\end{enumerate}
Putting everything together, we obtain
\begin{align}
  \label{res-Mt}
\widetilde M_{n,k}(t) &= \widetilde M_{n,k}^{(0)}\,
\left\{1-\frac{3m^2g_A^2}{(4\pi F)^2}
\left[\log\frac{m^2}{\mu^2}+1\right] \right\}
-\widetilde E_{n,k}^{(0)}\; \frac{m^2g_A^2}{(4\pi F)^2}\,
\log\frac{m^2}{\mu^2}
\nonumber \\[0.4em]
& \quad {}+\widetilde{M}_{n,k}^{(2,m)} m^2
      +\widetilde{M}_{n,k}^{(2,t)} t +O(q^3) \,,
\end{align}
where the terms going with $m^2$ and $t$ are due to tree-level
insertions.  With \eqref{res-Et}, \eqref{gpd-mom} and
\eqref{ff-inverse} one can write for the isoscalar quark GPD
$\widetilde E(x,\xi,t)$
\begin{align}
\widetilde E(x,\xi,t) &= \widetilde E^{(0)}(x,\xi)\,
\left\{1-\frac{3m^2g_A^2}{(4\pi F)^2}
\left[\log\frac{m^2}{\mu^2}+1\right] \right\}
-\widetilde H^{(0)}(x,\xi)\, \frac{m^2g_A^2}{(4\pi F)^2}\,
\log\frac{m^2}{\mu^2}
\nonumber \\[0.4em]
& \quad {}+m^2\slim\widetilde{E}^{(2,m)}(x,\xi)
      +t\slim \widetilde{E}^{(2,t)}(x,\xi) +O(q^3) \,.
\end{align}

\subsection{Vector operators}

The analysis of the vector operators proceeds along similar lines. The
operator $\mathcal{O}_{n,k,0}$ reads
\begin{equation}\label{Otnk}
\mathcal{O}_{n,k,0}(a)= b_{n,k}\, (ia\partial)^k\, \Nvbar\, N_v
+c_{n,k}\, (ia\partial)^{k-1}\, (i\partial_\mu)\,
\Nvbar\, [(S a),S^\mu] N_v + \ldots \,,
\end{equation}
where the $\ldots$ denote operators with fewer total derivatives.  One
finds $E_{n,k}^{(0)}= b_{n,k}$ and $M_{n,k-1}^{(0)}= -c_{n,k}$ for the
leading-order tree-level insertions, which implies $b_{n,k}=
c_{n,k+1}=0$ for odd $k$.

According to \eqref{OT} the graphs that give chiral corrections to
$E_{n,k}$ or $M_{n,k-1}$ must produce $k$ factors of $\Delta_\mu$
contracted with $a^\mu$ or $S^\mu$.  With the constraints from parity
invariance one finds that the operator $\mathcal{O}_{n,k,-1}$ does not
contain terms which have $k-1$ or more total derivatives contracted
with $a^\mu$ or $S^\mu$.  With the results of Section~\ref{sec:loop}
this implies that $E_{n,k}$ and $M_{n,k-1}$ do not receive corrections
from pion-nucleon operator insertions at order $O(q)$, and that
corresponding corrections at order $O(q^2)$ can only come from the
diagram in Fig.~\ref{fig-1}a with LO vertices.  One finds that the
one-loop contribution to the form factor $E_{n,k}$ is canceled by the
terms proportional to $g_A^2$ in the wave function renormalization
constant \eqref{ZN}.  For the form factor $M_{n,k}$ one obtains a
correction
\begin{equation}\label{res-M}
M_{n,k}^{(0)}
\left\{1-\frac{3m^2g_A^2}{(4\pi F)^2}\,\log\frac{m^2}{\mu^2}\right\}
\,.
\end{equation}
We note that for $n=1$, $k=0$ this implies a chiral logarithm for the
isoscalar magnetic form factor $G_{M,s}(t)$,
\begin{equation}
  \label{res-GM}
G_{M,s}(t) = \mu_s^{(0)}
  \left\{1-\frac{3m^2g_A^2}{(4\pi F)^2}\,\log\frac{m^2}{\mu^2}\right\}
  + G_{M,s}^{(2,m)} m^2 + G_{M,s}^{(2,t)} t + O(q^3) \,,
\end{equation}
where $\mu_s^{(0)}$ is the isoscalar magnetic moment of the nucleon in
the chiral limit and where we have added analytic terms due to
tree-level insertions.  The form \eqref{res-GM} is consistent with
the result of the relativistic calculation in \cite{Kubis:2000zd}.

\begin{figure}[t]
\psfrag{a}[c][c][0.9]{$wv-\frac{\Delta}{2}$}
\psfrag{c}[c][c][0.9]{$wv+\frac{\Delta}{2}$}
\psfrag{e}[c][c][0.9]{$l+wv$}
\psfrag{b}[c][c][0.9]{$l+\frac{\Delta}{2}$}
\psfrag{d}[c][c][0.9]{$l-\frac{\Delta}{2}$}
\psfrag{aa}[b][b][1.0]{a}
\psfrag{bb}[b][b][1.0]{b}
\centerline{\includegraphics[width=12cm]{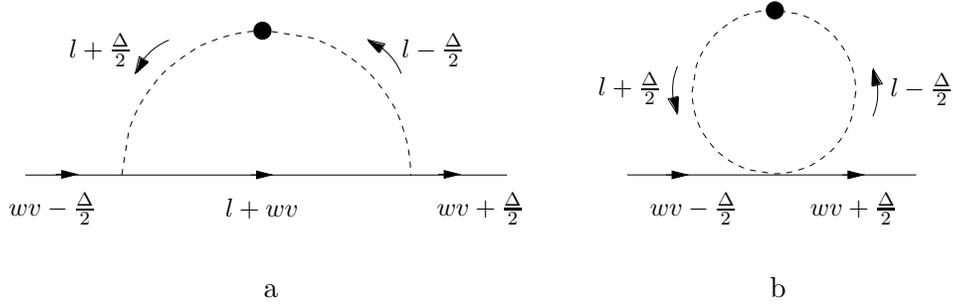}}
\caption{\label{fig-2} One-loop graphs with the insertion of the pion
operator ${\mathcal O}_{\pi}^n(a)$, denoted by a black blob.}
\end{figure}

One finally has to evaluate corrections due to the diagrams in
Fig.~\ref{fig-2} with insertion of the pion operator ${\mathcal
O}_{\pi}^n(a)$, where $n$ is even.  We use the representation of this
operator given in~\cite{DMS},\footnote{%
Note that the normalization of the twist-two operators
\protect\eqref{operators} used here differs from that in
\protect\cite{DMS} by a factor of 2.  The coefficients
$\tilde{a}_{n,k}$ have the same normalization here and in
\protect\cite{DMS}.}
\begin{equation}\label{ss-n}
{\cal O}_\pi^{n}(a)= F^2\,
\sum_{\substack{k=0\\ \mathrm{even}}}^{n-2}
\tilde a_{n,k}\, (i a\partial)^k \Tr \left[
(aL)\,(2i a \lrpartial)^{n-k-2} (aL)+
(aR)\,(2i a \lrpartial)^{n-k-2} (aR)
\right]
\end{equation}
with $L_\mu= U^\dagger\slim \partial_\mu U$ and $R_\mu = U
\partial_\mu U^\dagger$.
As discussed in Section~\ref{sec:loop}, the corrections to $M_{n,k}$
start at order $O(q^2)$ for $k=n-2$ and at higher order otherwise.
They are due to diagrams with LO vertices, so that only the graph in
Fig.~\ref{fig-2}a contributes.  This is because the leading-order
$\pi\pi NN$ vertex corresponds to an isovector transition of the
nucleon, as follows from \eqref{L2} and the expansion of $\Gamma$ in
\eqref{VA}.  Combining the result with the correction in \eqref{res-M}
and adding analytic terms from tree-level insertions, we obtain
\begin{equation}
  \label{res-full-M}
M_{n,k}(t) = M_{n,k}^{(0)}
\left\{1-\frac{3m^2g_A^2}{(4\pi F)^2}\,\log\frac{m^2}{\mu^2}\right\}
+ \delta_{k,n-2}^{\phantom{0}}\, M_{n}^{(2,\pi)}(t)
+ M_{n,k}^{(2,m)} m^2 + M_{n,k}^{(2,t)} t + O(q^3) \,,
\end{equation}
where $k$ is even and
\begin{align}
  \label{result-M2}
M_{n}^{(2,\pi)}(t) &=
\frac{3g_A^2}{(4\pi F)^2}\,
\sum_{\substack{j=0\\ \mathrm{even}}}^{n-2} {\tilde a}_{n,n-j-2}
\int_{-1}^1d\eta
\left[ \frac{\partial^2}{\partial\eta^2}\, \eta^j(1-\eta^2) \right]
m^2(\eta)\, \log\frac{m^2(\eta)}{\mu^2}
\nonumber\\[2mm]
&= \frac{3g_A^2}{(4\pi F)^2}\,
\sum_{\substack{j=2\\ \mathrm{even}}}^{n} 2^{-j} j(j-1)\,
A_{n,n-j}^{\pi\slim (0)}
\int_{-1}^1d\eta\,\eta^{j-2}\,m^2(\eta)\,\log\frac{m^2(\eta)}{\mu^2}
\end{align}
with
\begin{equation}
m^2(\eta)=m^2- \frac{t}{4}\slim (1-\eta^2) \,.
\end{equation}
Here $A_{n,k}^{\pi\slim (0)}$ is the chiral limit of the form factors
$A_{n,k}^\pi(t)$ describing the moments of the pion isoscalar GPD,
\begin{align}
\int_{-1}^1 dx\, x^{n-1} H_\pi^{I=0}(x,\xi,t)
=\sum_{\substack{k=0\\ \mathrm{even}}}^n (2\xi)^k A_{n,k}^\pi(t) .
\end{align}
The relation to the low-energy constants $\tilde a_{n,k}$
reads~\cite{DMS}
\begin{align}
A_{n,k}^{\pi\slim (0)}=2^{n-k}\,
  \Big[\tilde a_{n,k-2}-\tilde a_{n,k}\Big]\,,
\end{align}
which implies
\begin{align}
\tilde{a}_{n,n-k} &= {}- \sum_{\substack{j=k\\ \mathrm{even}}}^n
  2^{-j} A_{n,n-j}^{\pi\slim (0)} \hspace{2em}\text{for~} k>0 \,, &
\sum_{\substack{j=0\\ \mathrm{even}}}^n 2^{-j}
  A_{n,n-j}^{\pi\slim (0)} &= 0 \,.
\end{align}
The corrections to $E_{n,k}$ start at order $O(q)$ for $k=n$ and at
order $O(q^3)$ or higher otherwise.  At one-loop order we obtain $O(q)$
corrections to $E_{n,n}$ from graphs involving only LO vertices.
Corrections of order $O(q^2)$ involve either graphs with one NLO
vertex or propagator correction, or graphs with LO vertices and the
subleading part $wv$ of the residual nucleon momenta, see the
discussion after \eqref{pow-count}.  Our final result including
analytic terms from tree-level insertions is
\begin{align}\label{res-E}
E_{n,k}(t)=E_{n,k}^{(0)}
+\delta_{n,k}^{\phantom{0}}
  \Bigl[E_{n}^{(1,\pi)}(t)+E_{n}^{(2,\pi)}(t)\Bigr]
+ E_{n,k}^{(2,m)} m^2 + E_{n,k}^{(2,t)} t +O(q^3)\,,
\end{align}
where the order $O(q)$ correction reads
\begin{align}
  \label{result-E1}
E_{n}^{(1,\pi)}(t) &= {}-M (2m^2-t)\,
\frac{3\pi g_A^2}{8\slim (4\pi F)^2}\,
\sum_{\substack{j=0\\ \mathrm{even}}}^{n-2} {\tilde a}_{n,n-j-2}\,
\int_{-1}^1d\eta\, \frac{\eta^j (1-\eta^2)}{m(\eta)}
\nonumber\\[2mm]
&= M (2m^2-t)\,\frac{3\pi g_A^2}{8\slim (4\pi F)^2}\,
\sum_{\substack{j=2\\ \mathrm{even}}}^{n}
2^{-j}\,A_{n,n-j}^{\pi\slim (0)}\int_{-1}^1d\eta\,
\frac{1-\eta^j}{m(\eta)}\,,
\end{align}
and the order $O(q^2)$ term is
\begin{align}
  \label{result-E2}
E_{n}^{(2,\pi)}(t) =& ~\frac{3m^2 g_A^2}{(4\pi F)^2}\,
\log\frac{m^2}{\mu^2}\;
\sum_{\substack{j=0\\ \mathrm{even}}}^{n-2} \tilde a_{n,n-j-2}
\nonumber\\[2mm]
&
+\frac{6}{(4\pi F)^2}\,
\sum_{\substack{j=0\\ \mathrm{even}}}^{n-2} \tilde a_{n,n-j-2}
\int_{-1}^{1}d\eta\, \eta^j (1-\eta^2)\,
\Biggl\{ \frac{g_A^2}{32}
 \left[2t \left(\log\frac{m^2(\eta)}{\mu^2}+1\right)
 -\frac{(t-2m^2)^2}{m^2(\eta)} \,\right]
\nonumber\\[2mm]
& ~~+ M \left[c_1\slim m^2 \left(\log\frac{m^2(\eta)}{\mu^2}+1\right)
-\frac{3}{4} c_2\slim m^2(\eta)\, \log\frac{m^2(\eta)}{\mu^2}
-c_3\slim m^2(\eta)
  \left(\log\,\frac{m^2(\eta)}{\mu^2}+\frac{1}{2}\right)
\right] \Biggr\}
\nonumber\\[2mm]
=& {}- \frac{3m^2 g_A^2}{2\slim (4\pi F)^2}\,
\log\frac{m^2}{\mu^2}\,
\sum_{\substack{j=2\\ \mathrm{even}}}^{n}2^{-j} j\,
A_{n,n-j}^{\pi\slim (0)}
\nonumber\\[2mm]
& -\frac{6}{(4\pi F)^2}\, \sum_{\substack{j=2\\ \mathrm{even}}}^{n}
2^{-j}\,A_{n,n-j}^{\pi\slim (0)} \int_{-1}^1d\eta\, (1-\eta^j)\,
\Biggl\{
\frac{g_A^2}{32}\left[2t \left(\log\,\frac{m^2(\eta)}{\mu^2}+1\right)
-\frac{(t-2m^2)^2}{m^2(\eta)} \,\right]
\nonumber\\[2mm]
& ~~+ M \left[c_1\slim m^2 \left(\log\frac{m^2(\eta)}{\mu^2}+1\right)
-\frac{3}{4}c_2\slim m^2(\eta)\, \log\frac{m^2(\eta)}{\mu^2}
-c_3\slim m^2(\eta)
  \left(\log\frac{m^2(\eta)}{\mu^2}+\frac{1}{2}\right)
\right] \Biggr\}
\nonumber \\
\end{align}
The integrals over $\eta$ in \eqref{result-M2}, \eqref{result-E1} and
\eqref{result-E2} are elementary, but we have not found a simple
closed form of the result for general $n$.  In the next section we
give explicit results for the values and derivatives at $t=0$ of the
form factors.

Our result for the form factor $\widetilde M_{n,k}(t)$ disagrees
with~\cite{ando}, where it was taken for granted that the only
operators which contribute at order $O(q^2)$ are those which already
appear at tree-level in the same form factor.  As our analysis shows,
this holds indeed in many cases but not in all.  For all other form
factors our results agree with \cite{ando} where
comparable.\footnote{%
Note that \protect\cite{ando} gives the correction to $E_{n,n}$ at
order $O(q)$ but not at order $O(q^2)$.}
For $n=2$ our results for the vector operators also agree with those
of Belitsky and Ji~\cite{BJi}.\footnote{%
When comparing results, one must take into account that
\protect\cite{BJi} uses $\overline{\mathrm{MS}}$ renormalization,
where for each pole in $4-2\epsilon$ dimensions one subtracts
$1/\epsilon+\log(4\pi)+\psi(1)$, whereas we use the scheme of
\protect\cite{ChPT} and subtract $1/\epsilon+\log(4\pi)+\psi(2)$.}


\section{Results for moments of GPDs}
\label{sec:gpd-res}

We now transform the results of the previous section to the basis of
the form factors $A_{n,k}$, $B_{n,k}$, $C_n$ and
$\widetilde{A}_{n,k}$, $\widetilde{B}_{n,k}$ corresponding to moments
of GPDs in the conventional parameterization.  We give the values and
derivatives of these form factors at $t=0$, which allows us to obtain
closed expressions.  Furthermore, these quantities are of most
immediate interest in studies of GPDs on the lattice.

With our results \eqref{res-Et}, \eqref{res-Mt}, \eqref{res-full-M},
\eqref{res-E} and the conversion formulae \eqref{ff-trafo},
\eqref{ff-inverse} one obtains for the form factors at $t=0$
\begin{align}
  \label{res-gpd-0}
\widetilde{A}_{n,k}(0) &= \widetilde{A}_{n,k}^{(0)}\,
  \left\{1-\frac{3m^2g_A^2}{(4\pi F)^2}
  \left[\log\frac{m^2}{\mu^2}+1\right] \right\}
  + \widetilde{A}_{n,k}^{(2,m)} m^2 + O(m^3) \,,
\nonumber \\[0.1em]
\widetilde{B}_{n,k}(0) &= \widetilde{B}_{n,k}^{(0)}\,
  \left\{1-\frac{3m^2g_A^2}{(4\pi F)^2}
  \left[\log\frac{m^2}{\mu^2}+1\right] \right\}
  - \widetilde{A}_{n,k}^{(0)}\; \frac{m^2g_A^2}{(4\pi F)^2}\,
    \log\frac{m^2}{\mu^2}
  + \widetilde{B}_{n,k}^{(2,m)} m^2 + O(m^3) \,,
\nonumber \\[0.3em]
A_{n,k}(0) &= A_{n,k}^{(0)} + A_{n,k}^{(2,m)} m^2 + O(m^3) \,,
\nonumber \\[0.3em]
B_{n,k}(0) &= B_{n,k}^{(0)} - \bigl( A_{n,k}^{(0)}
   +B_{n,k}^{(0)}\bigr)\, \frac{3m^2g_A^2}{(4\pi F)^2}\,
     \log\frac{m^2}{\mu^2}
   +\delta_{k,n-2}^{\phantom{0}}\, M_{n}^{(2,\pi)}(0)
   + B_{n,k}^{(2,m)} m^2 + O(m^3) \,,
\nonumber \\[0.4em]
C_{n}(0) &= C_{n}^{(0)} + E_{n}^{(1,\pi)}(0) + E_{n}^{(2,\pi)}(0)
  + C_{n}^{(2,m)} m^2 + O(m^3)
\end{align}
with coefficients related to those in Section~\ref{sec:results} by
$\widetilde{A}_{n,k}^{(0)}= \widetilde{E}_{n,k}^{(0)}$,
$\widetilde{B}_{n,k}^{(0)}= \widetilde{M}_{n,k}^{(0)} -\half
\widetilde{E}_{n,k}^{(0)}$, $A_{n,k}^{(0)}= E_{n,k}^{(0)}$,
$B_{n,k}^{(0)}= M_{n,k}^{(0)}- E_{n,k}^{(0)}$,
$C_{n\phantom{l}}^{(0)}= E_{n,n\phantom{l}}^{(0)}$ and by analogous
relations for the coefficients with superscript $(2,m)$.  Setting $m$,
$g_A$, $F$ to their physical values and choosing $\mu=M$, one
finds that the corrections from loop graphs with nucleon operator
insertions in \eqref{res-gpd-0} are moderately large, with $3 m^2
g_A^2\, (4\pi F)^{-2}\, [\slim \log(m^2/\mu^2)+1 \slim] \approx -0.20$
and $m^2 g_A^2\, (4\pi F)^{-2} \log(m^2/\mu^2) \approx -0.09$.  In the
case of $B_{n,k}$ this loop correction can be substantial if
$|B_{n,k}| \ll |A_{n,k}|$, which is empirically found for the
electromagnetic form factors (i.e.\ the case $n=1$) and also in
lattice evaluations \cite{QCDSF} for the moments with $n=2$.  The
contributions to $B_{n,n-2}(0)$ and $C_n(0)$ from loop graphs with
pion operator insertions are
\begin{align}
\label{pi-op-insertions}
M_{n}^{(2,\pi)}(0) &= \frac{6m^2 g_A^2}{(4\pi F)^2}\,
  \log\frac{m^2}{\mu^2}\,
  \sum_{\substack{j=2\\ \mathrm{even}}}^{n}
  2^{-j} j\, A_{n,n-j}^{\pi\slim (0)} \,,
\nonumber \\[0.1em]
E_{n}^{(1,\pi)}(0) &= \frac{3\pi\slim m M g_A^2}{2\slim (4\pi F)^2}\,
  \sum_{\substack{j=2\\ \mathrm{even}}}^{n}
  2^{-j}\, \frac{j}{j+1}\, A_{n,n-j}^{\pi\slim (0)} \,,
\nonumber \\[0.1em]
E_{n}^{(2,\pi)}(0) &= {}- \frac{3m^2 g_A^2}{2\slim (4\pi F)^2}\,
  \log\frac{m^2}{\mu^2}\,
  \sum_{\substack{j=2\\ \mathrm{even}}}^{n}
  2^{-j} j\, A_{n,n-j}^{\pi\slim (0)}
\nonumber \\[0.1em]
& \hspace{-3.3em}{}+ \frac{12 m^2}{(4\pi F)^2}\,
    \left\{ \frac{g_A^2}{8} - M \left[
      c_1 \left(\log\frac{m^2}{\mu^2}+1\right)
    - \frac{3}{4} c_2\slim \log\frac{m^2}{\mu^2}
    - c_3 \left(\log\frac{m^2}{\mu^2}+\frac{1}{2}\right)
    \right] \right\}
    \sum_{\substack{j=2\\ \mathrm{even}}}^{n}
    2^{-j} \frac{j}{j+1}\, A_{n,n-j}^{\pi\slim (0)} \,.
\nonumber \\
\end{align}
Setting $M$, $m$, $g_A$, $F$ to their physical values, choosing
$\mu=M$, and taking the estimates $c_1\approx -0.9\gev^{-1}$,
$c_2\approx 3.3\gev^{-1}$, $c_3\approx -4.7\gev^{-1}$ from
\cite{Meissner:2005ba} we find $M_{2}^{(2,\pi)}(0) \approx -0.27\,
A_{2,0}^{\pi\slim (0)}$ and $E_{2}^{(1,\pi)}(0)+E_{2}^{(2,\pi)}(0)
\approx (0.12 + 0.17)\, A_{2,0}^{\pi\slim (0)}$.  At the physical
point, the order $O(m)$ correction is hence not very large.  The full
size of the order $O(m^2)$ corrections depends of course on the
analytic terms in \eqref{res-gpd-0}, whose values are not known.

The derivatives of the form factors at $t=0$ obtain nonanalytic
contributions only from the pion operator insertions.  Writing
$\partial_t A(0) = \bigl[\frac{\partial}{\partial t} A(t)\bigr]_{\,
t=0}$ etc.\ we have
\begin{align}
\partial_t \widetilde{A}_{n,k}(0) &= \widetilde{E}_{n,k}^{(2,t)}
   + O(m) \,,
\nonumber \\[0.15em]
\partial_t \widetilde{B}_{n,k}(0) &= \widetilde{M}_{n,k}^{(2,t)}
  - \widetilde{E}_{n,k}^{(2,t)} \big/2
  - \widetilde{E}_{n,k}^{(0)} \slim\big/(32 M^2)
  + O(m) \,,
\nonumber \\[0.15em]
\partial_t A_{n,k}(0) &= E_{n,k}^{(2,t)}
  - \bigl( M_{n,k}^{(0)}-E_{n,k}^{(0)} \bigr) \big/(4 M^2) + O(m) \,,
\nonumber \\[0.15em]
\partial_t B_{n,k}(0) &=
  \delta_{k,n-2}\; \partial_t M_{n}^{(2,\pi)}(0)
  + M_{n,k}^{(2,t)}-E_{n,k}^{(2,t)}
  + \bigl( M_{n,k}^{(0)}-E_{n,k}^{(0)} \bigr) \big/(4 M^2) + O(m) \,,
\nonumber \\[0.3em]
\partial_t C_{n}(0) &=
  \partial_t E_{n}^{(1,\pi)}(0) + \partial_t E_{n}^{(2,\pi)}(0)
  + E_{n,n}^{(2,t)} + E_{n,n}^{(0)} \big/(4 M^2)
  + O(m)
\end{align}
with
\begin{align}
\label{pi-op-insertions-deriv}
\partial_t M_{n}^{(2,\pi)}(0) &= {}-\frac{3g_A^2}{(4\pi F)^2}
  \left[\log\frac{m^2}{\mu^2}+1\right]\,
  \sum_{\substack{j=2\\ \mathrm{even}}}^{n}
  2^{-j} \frac{j}{j+1}\, A_{n,n-j}^{\pi\slim (0)} \,,
\nonumber \\[0.1em]
\partial_t E_{n}^{(1,\pi)}(0) &= {}- \frac{M}{m}\,
  \frac{\pi g_A^2}{8\slim (4\pi F)^2}\,
  \sum_{\substack{j=2\\ \mathrm{even}}}^{n}
  2^{-j}\, \frac{j\, (5j+14)}{(j+1) (j+3)}\,
  A_{n,n-j}^{\pi\slim (0)} \,,
\nonumber \\[0.1em]
\partial_t E_{n}^{(2,\pi)}(0) &= {}- \frac{3g_A^2}{4\slim (4\pi F)^2}\,
  \left[\log\frac{m^2}{\mu^2} +3\right]\,
  \sum_{\substack{j=2\\ \mathrm{even}}}^{n}
  2^{-j} \frac{j}{j+1}\, A_{n,n-j}^{\pi\slim (0)}\,
\nonumber \\[0.1em]
 & \hspace{-4.2em}{}+ \frac{2}{(4\pi F)^2}\,
  \left\{ \frac{g_A^2}{8} + M \left[
      c_1 - \frac{3}{4} c_2 \left(\log\frac{m^2}{\mu^2}+1\right)
    - c_3 \left(\log\frac{m^2}{\mu^2}+\frac{3}{2}\right)
    \right] \right\}\,
  \sum_{\substack{j=2\\ \mathrm{even}}}^{n}
  2^{-j} \frac{j\, (j+4)}{(j+1) (j+3)}\, A_{n,n-j}^{\pi\slim (0)} \,.
\nonumber \\
\end{align}
Note that in the chiral limit the derivative $\partial_t B_{n,n-2}(0)
\sim \partial_t M_{n}^{(2,\pi)}(0)$ diverges as $\log(m^2 /\mu^2)$ and
$\partial_t C_{n}(0) \sim \partial_t E_{n}^{(1,\pi)}(0)$ as $1/m$.
With the parameters specified above, one finds $\partial_t
M_{2}^{(2,\pi)}(0) \approx 1.7\gev^{-2}\, A_{2,0}^{\pi\slim (0)}$ and
$\partial_t E_{2}^{(1,\pi)}(0)+\partial_t E_{2}^{(2,\pi)}(0) \approx
-(2.5 + 1.2)\gev^{-2}\, A_{2,0}^{\pi\slim (0)}$.  Numerically, the
term $\partial_t E_{n}^{(1,\pi)}(0)$ is thus important but not
extremely large at the physical point.


\section{Summary}
\label{sec:sum}

Using heavy-baryon chiral perturbation theory, we have calculated the
chiral corrections up to order $O(q^2)$ for the form factors which
parameterize moments of nucleon GPDs.  We have restricted ourselves to
vector and axial-vector quark distributions in the isosinglet
combination.  Our results generalize trivially to the corresponding
gluon GPDs, which have the same quantum numbers and therefore the same
corresponding operators in the effective theory (except for the values
of the matching constants).  Our method is also applicable to
operators of different tensor or flavor structure.

The moments of GPDs contain terms of different order in $1/M$, ranging
from $M^{n-1}$ to $M^{-1}$. We have shown that, due to the way in
which factors $v_\mu$ and $\Delta_\mu$ arise in the calculation, the
number of loops and the order in the expansion of the effective
Lagrangian required to calculate a form factor to a given order
$O(q^{\slim d})$ does not grow with the number of factors $\Delta_\mu$
that accompany the form factor in the nucleon matrix element.  A
general power-counting formula is given after \eqref{pow-count}.  In
the case of the form factors $\widetilde{M}_{n,k}(t)$, calculation of
the order $O(q^2)$ correction requires the pion-nucleon Lagrangian up
to third order.

We have found that the form factors $\widetilde{E}_{n,k}(t)$ and
$\widetilde{M}_{n,k}(t)$ receive corrections of order $O(q^2)$ which,
apart from analytic terms, are independent of the moment indices and
independent of $t$.  The same holds for the one-loop corrections to
$M_{n,k}(t)$ with $k<n-2$, whereas the corresponding corrections for
$E_{n,k}$ with $k<n$ are zero.  The form factors $M_{n,n-2}$ receive
additional corrections at order $O(q^2)$ from one-loop graphs with the
insertion of pion operators, and $E_{n,n}$ receives corresponding
contributions starting at order $O(q)$.

For the form factors parameterizing moments of isoscalar GPDs, we find
that $B_{n,k}$, $\widetilde{A}_{n,k}$ and $\widetilde{B}_{n,k}$ at
$t=0$ receive nonanalytic corrections of the form $m^2
\log(m^2/\mu^2)$ from loops with nucleon operator insertions.  No such
corrections are found for $A_{n,k}$ and $C_{n}$.  The form factors
$B_{n,n-2}$ at $t=0$ receive in addition $m^2 \log(m^2/\mu^2)$
corrections from loop graphs with pion operator insertions, and the
corresponding nonanalytic contributions to $C_{n}$ give a term
proportional to $m$.  To leading chiral order, loop graphs with pion
operator insertions are the only source of nonanalytic $m^2$
dependence for the derivatives of the form factors at $t=0$.  The
derivative of $M_{n,n-2}$ diverges like $\log(m^2/\mu^2)$ in the
chiral limit, and the derivative of $C_{n}$ like $1/m$.


\section*{Acknowledgments}

We are grateful to U.-G.\ Mei{\ss}ner for clarifying discussions and
valuable remarks on the manuscript.  This work is supported by the
Helmholtz Association, contract number VH-NG-004.  A.S.\ acknowledges
the hospitality of INT, Seattle, USA (program INT-06-1: ``Exploration
of Hadron Structure and Spectroscopy using Lattice QCD''), where part
of this work was done.


\section*{Erratum}

The result for $E_n^{(2,\pi)}(t)$ in \eqref{result-E2} contains mistakes in the terms that multiply the low-energy constants $c_2$ and $c_3$.  The corrected version of this equation reads
\begin{align*}
E_{n}^{(2,\pi)}(t) =& ~\frac{3m^2 g_A^2}{(4\pi F)^2}\,
\log\frac{m^2}{\mu^2}\;
\sum_{\substack{j=0\\ \mathrm{even}}}^{n-2} \tilde a_{n,n-j-2}
\nonumber\\[2mm]
&
+\frac{6}{(4\pi F)^2}\,
\sum_{\substack{j=0\\ \mathrm{even}}}^{n-2} \tilde a_{n,n-j-2}
\int_{-1}^{1}d\eta\, \eta^j (1-\eta^2)\,
\Biggl\{ \frac{g_A^2}{32}
 \left[2t \left(\log\frac{m^2(\eta)}{\mu^2}+1\right)
 -\frac{(t-2m^2)^2}{m^2(\eta)} \,\right]
\nonumber\\[2mm]
& ~~+ M
\left[\left(c_1m^2+\textcolor{red}{\frac14c_3\left(t-2m^2\right)}\right) \left(\log\frac{m^2(\eta)}{\mu^2}+1\right)
-\textcolor{red}{ \frac{1}{4}c_2 } \slim m^2(\eta)\, \log\frac{m^2(\eta)}{\mu^2}
\right]
\Biggr\}
\nonumber\\[2mm]
=& {}- \frac{3m^2 g_A^2}{2\slim (4\pi F)^2}\,
\log\frac{m^2}{\mu^2}\,
\sum_{\substack{j=2\\ \mathrm{even}}}^{n}2^{-j} j\,
A_{n,n-j}^{\pi\slim (0)}
\nonumber\\[2mm]
& -\frac{6}{(4\pi F)^2}\, \sum_{\substack{j=2\\ \mathrm{even}}}^{n}
2^{-j}\,A_{n,n-j}^{\pi\slim (0)} \int_{-1}^1d\eta\, (1-\eta^j)\,
\Biggl\{
\frac{g_A^2}{32}\left[2t \left(\log\,\frac{m^2(\eta)}{\mu^2}+1\right)
-\frac{(t-2m^2)^2}{m^2(\eta)} \,\right]
\nonumber\\[2mm]
& ~~+ M \left[\left(c_1m^2+\textcolor{red}{\frac14c_3\left(t-2m^2\right)}\right) \left(\log\frac{m^2(\eta)}{\mu^2}+1\right)
-\textcolor{red}{ \frac{1}{4}c_2 } \slim m^2(\eta)\, \log\frac{m^2(\eta)}{\mu^2}
\right] \Biggr\} \,,
\end{align*}
where the terms that have changed are marked in red.  The corresponding corrections in equations \eqref{pi-op-insertions} and \eqref{pi-op-insertions-deriv} read
\begin{align*}
E_{n}^{(2,\pi)}(0) &= {}- \frac{3m^2 g_A^2}{2\slim (4\pi F)^2}\,
  \log\frac{m^2}{\mu^2}\,
  \sum_{\substack{j=2\\ \mathrm{even}}}^{n}
  2^{-j} j\, A_{n,n-j}^{\pi\slim (0)}
\nonumber \\[0.1em]
& \hspace{-3.3em}{}+ \frac{12 m^2}{(4\pi F)^2}\,
    \left\{ \frac{g_A^2}{8} - M \left[
      \left(c_1 \textcolor{red}{-\frac12{c_3}}\right)
      \left(\log\frac{m^2}{\mu^2}+1\right)
    - \textcolor{red}{\frac{1}{4} c_2 }\slim \log\frac{m^2}{\mu^2}
    \right] \right\} \,
    \sum_{\substack{j=2\\ \mathrm{even}}}^{n}
    2^{-j} \frac{j}{j+1}\, A_{n,n-j}^{\pi\slim (0)}
\end{align*}
and
\begin{align*}
\partial_t E_{n}^{(2,\pi)}(0) &= {}- \frac{3g_A^2}{4\slim (4\pi F)^2}\,
  \left[\log\frac{m^2}{\mu^2} +3\right]\,
  \sum_{\substack{j=2\\ \mathrm{even}}}^{n}
  2^{-j} \frac{j}{j+1}\, A_{n,n-j}^{\pi\slim (0)}
  + \frac{2}{(4\pi F)^2}
\nonumber \\[0.1em]
 & \hspace{-4.2em}{} \times
   \sum_{\substack{j=2\\ \mathrm{even}}}^{n} \,
   \left\{ \frac{g_A^2}{8} + M \left[
      c_1 \textcolor{red}{ - \frac12{c_3}- \left(\frac{1}{4} c_2 +\frac{3(j+3)}{2(j+4)}\slim c_3\right) } \left(\log\frac{m^2}{\mu^2}+1\right)
    \right] \right\} \,
    2^{-j} \frac{j\, (j+4)}{(j+1) (j+3)}\, A_{n,n-j}^{\pi\slim (0)} \,,
\end{align*}
respectively.  With the parameter estimates $c_1\approx -0.9\gev^{-1}$, $c_2\approx 3.3\gev^{-1}$, $c_3\approx -4.7\gev^{-1}$ given below \eqref{pi-op-insertions}, one obtains corrected numerical values
\begin{align*}
E_{2}^{(1,\pi)}(0)+E_{2}^{(2,\pi)}(0)
 & \approx (0.12 + \textcolor{red}{0.10})\, A_{2,0}^{\pi\slim (0)} \,,
\\[0.2em]
\partial_t E_{2}^{(1,\pi)}(0)+\partial_t E_{2}^{(2,\pi)}(0)
 & \approx -(2.5 + \textcolor{red}{3.4})\gev^{-2}\, A_{2,0}^{\pi\slim (0)} \,.
\end{align*}
After the above corrections, we agree with the results obtained in \cite{Alharazin:2020yjv}.  We thank M.~Polyakov for pointing us to the difference between the calculation in that paper and the published version of our work.



\begin{thebibliography}{99}

\bibitem{GPDs}
D.~M{\"u}ller, D.~Robaschik, B.~Geyer, F.~M. Dittes and
J.~Ho\v{r}ej\v{s}i,
Fortschr. Phys. {\bf 42} (1994) 101 [hep-ph/9812448].

\bibitem{Ji}
X.~D.~Ji,
Phys. Rev. Lett. {\bf 78} (1997) 610 [hep-ph/9603249].

\bibitem{R97}
A.~V. Radyushkin,
Phys. Rev. {\bf D56} (1997) 5524 [hep-ph/9704207].

\bibitem{Ji98}
  X.~D.~Ji,
  J.\ Phys.\ G {\bf 24} (1998) 1181
  [hep-ph/9807358].

\bibitem{GPV}
K.~Goeke, M.~V.~Polyakov and M.~Vanderhaeghen,
Prog.\ Part.\ Nucl.\ Phys.\  {\bf 47} (2001) 401 [hep-ph/0106012].

\bibitem{Diehl03}
M.~Diehl,
Phys.\ Rept.\  {\bf 388} (2003) 41 [hep-ph/0307382].

\bibitem{Belitsky:2005qn}
A.~V.~Belitsky and A.~V.~Radyushkin,
Phys.\ Rept.\  {\bf 418} (2005) 1 [hep-ph/0504030].

\bibitem{blumlein}
J.~Bl\"umlein, B.~Geyer and D.~Robaschik,
Phys.\ Lett.\ B {\bf 406} (1997) 161 [hep-ph/9705264].

\bibitem{QCDSF}
M.~G\"ockeler {\it et al.} [QCDSF Collab.],
Phys.\ Rev.\ Lett.\  {\bf 92} (2004) 042002 [hep-ph/0304249];
%
Phys.\ Lett.\ B {\bf 627} (2005) 113
[hep-lat/0507001];\\
%
Ph.~H\"agler  {\it et al.} [LHPC Collab.],
Phys.\ Rev.\ D {\bf 68} (2003) 034505 [hep-lat/0304018];
%
Phys.\ Rev.\ Lett.\  {\bf 93} (2004) 112001 [hep-lat/0312014].

\bibitem{exp}
A.~Airapetian {\it et al.}  [HERMES Collab.],
Phys.\ Rev.\ Lett.\  {\bf 87} (2001) 182001
[hep-ex/0106068];\\
%
A.~Aktas {\it et al.}  [H1 Collab.],
Eur.\ Phys.\ J.\ C {\bf 44} (2005) 1
[hep-ex/0505061];\\
%
S.~Chekanov {\it et~al.} [ZEUS Collab.],
Phys.\ Lett.\ B {\bf 573} (2003) 46
[hep-ex/0305028];\\
%
S.~Stepanyan {\it et~al.} [CLAS Collab.],
Phys. Rev. Lett. {\bf 87} (2001) 182002
[hep-ex/0107043].

\bibitem{Burkardt:2002hr}
M.~Burkardt,
Int.\ J.\ Mod.\ Phys.\ A {\bf 18} (2003) 173
[hep-ph/0207047];\\
%
M.~Diehl and Ph.~H\"agler,
Eur.\ Phys.\ J.\ C {\bf 44} (2005) 87
[hep-ph/0504175].

\bibitem{Polyakov:2002yz}
M.~V.~Polyakov,
Phys.\ Lett.\ B {\bf 555} (2003) 57
[hep-ph/0210165].

\bibitem{Belitsky:2003nz}
A.~V.~Belitsky, X.~D.~Ji and F.~Yuan,
Phys.\ Rev.\ D {\bf 69} (2004) 074014
[hep-ph/0307383].

\bibitem{ChPT}
J.~Gasser and H.~Leutwyler,
Annals Phys.\  {\bf 158} (1984) 142.

\bibitem{chvol}
A.~Ali Khan {\it et al.}  [QCDSF-UKQCD Collab.],
  Nucl.\ Phys.\ B {\bf 689} (2004) 175
  [hep-lat/0312030];\\
%
M.~G\"ockeler {\it et al.} [QCDSF Collab.],
  Phys.\ Rev.\ D {\bf 71} (2005) 034508
  [hep-lat/0303019].

\bibitem{baer}
O.~B\"ar,
  Nucl.\ Phys.\ Proc.\ Suppl.\  {\bf 140} (2005) 106
  [hep-lat/0409123].

\bibitem{CJi}
  J.~W.~Chen and X.~D.~Ji,
  Phys.\ Rev.\ Lett.\  {\bf 88} (2002) 052003
  [hep-ph/0111048].

\bibitem{KP}
N.~Kivel and M.~V.~Polyakov,
hep-ph/0203264.

\bibitem{DMS}
  M.~Diehl, A.~Manashov and A.~Sch\"afer,
  Phys.\ Lett.\ B {\bf 622} (2005) 69
  [hep-ph/0505269].

\bibitem{BFHM}
  V.~Bernard, H.~W.~Fearing, T.~R.~Hemmert and U.-G.~Mei{\ss}ner,
  Nucl.\ Phys.\ A {\bf 635} (1998) 121,
  Erratum ibid.\ A {\bf 642} (1998) 563
  [hep-ph/9801297].

\bibitem{BJi}
  A.~V.~Belitsky and X.~D.~Ji,
  Phys.\ Lett.\ B {\bf 538} (2002) 289
  [hep-ph/0203276].

\bibitem{JM}
  E.~Jenkins and A.~V.~Manohar,
  Phys.\ Lett.\ B {\bf 255} (1991) 558.

\bibitem{B92}
  V.~Bernard, N.~Kaiser, J.~Kambor and U.-G.~Mei{\ss}ner,
  Nucl.\ Phys.\ B {\bf 388} (1992) 315.

\bibitem{BKM}
  V.~Bernard, N.~Kaiser and U.-G.~Mei{\ss}ner,
  Int.\ J.\ Mod.\ Phys.\ E {\bf 4} (1995) 193
  [hep-ph/9501384].

\bibitem{Meissner:2005ba}
U.-G.~Mei{\ss}ner,
PoS {\bf LAT2005} (2006) 009
[hep-lat/0509029].

\bibitem{Steininger:1998ya}
S.~Steininger, U.-G.~Mei{\ss}ner and N.~Fettes,
JHEP {\bf 9809} (1998) 008
[hep-ph/9808280];\\
J.~Kambor and M.~Moj\v{z}i\v{s},
JHEP {\bf 9904} (1999) 031
[hep-ph/9901235].

\bibitem{Fettes:1998ud}
N.~Fettes, U.-G.~Mei{\ss}ner and S.~Steininger,
Nucl.\ Phys.\ A {\bf 640} (1998) 199
[hep-ph/9803266].

\bibitem{ASv}
  D.~Arndt and M.~J.~Savage,
  Nucl.\ Phys.\ A {\bf 697} (2002) 429
  [nucl-th/0105045].

\bibitem{Kubis:2000zd}
B.~Kubis and U.-G.~Mei{\ss}ner,
Nucl.\ Phys.\ A {\bf 679} (2001) 698
[hep-ph/0007056].

\bibitem{ando}
S.~I.~Ando, J.~W.~Chen and C.~W.~Kao,
Phys.\ Rev.\ D \textbf{74} (2006) 094013
[hep-ph/0602200].

\bibitem{Alharazin:2020yjv}
H.~Alharazin, D.~Djukanovic, J.~Gegelia and M.~Polyakov,
arXiv:2006.05890 [hep-ph].

\end{thebibliography}
\end{document}